\documentclass[MSNbibl,nameyear,dvips]{arxstspdf}
\usepackage{dcolumn}
\usepackage{multirow}
\usepackage{graphicx}
\usepackage{flushend}
\usepackage{stfloats}

\volume{28}
\issue{2}
\pubyear{2013}
\firstpage{189}
\lastpage{208}
\doi{10.1214/12-STS406} %kopijuoti is PTS
\makeatletter

\newcolumntype{k}[1]{D{,}{}{#1}}
\newcolumntype{d}[1]{D{.}{.}{#1}}

\newcommand{\tnote}{\tabnoteref}

\setattribute{abstract}{width}{365pt}
\setattribute{keyword}{width}{365pt}

\makeatother

\begin{document}
\begin{frontmatter}

\title{A Comparative Review of Dimension Reduction Methods in
Approximate Bayesian Computation}%\thanksref{T1}
% kai straipsnis turi susijusiu diskusiju ir rejoinder'iu
%rejoinder at \relateddoi{r}{10.1214/00-STSXXXX}.}
\runtitle{Dimension reduction methods in ABC}

\begin{aug}
\author[a]{\fnms{M. G. B.} \snm{Blum}\corref{}\ead[label=e1]{michael.blum@imag.fr}},
\author[b]{\fnms{M. A.} \snm{Nunes}},
\author[b]{\fnms{D.} \snm{Prangle}}
\and
\author[c]{\fnms{S. A.} \snm{Sisson}}
\runauthor{Blum, Nunes, Prangle and Sisson}

\affiliation{Universit\'e Joseph Fourier, Lancaster University,
Lancaster University and University of New South Wales}

\address[a]{M. G. B. Blum is Research Associate, Universit\'e Joseph Fourier, Centre
National de la Recherche Scientifique, Laboratoire TIMC-IMAG UMR 5525,
Grenoble, F-38041, France \printead{e1}.}
\address[b]{M. A. Nunes is Senior Research Associate and D. Prangle is Lecturer, Mathematics
and Statistics Department, Fylde College, Lancaster University,
Lancaster LA1 4YF, United
Kingdom.}
\address[c]{S. A. Sisson is Associate Professor, School of Mathematics and
Statistics, University of New South Wales, Sydney 2052,
Australia.}

\end{aug}

% ABSTRACT
%
\begin{abstract}
Approximate Bayesian computation (ABC) methods make use of comparisons
between simulated and observed summary statistics to overcome the
problem of computationally intractable likelihood functions. As the
practical implementation of ABC requires computations based on vectors
of summary statistics, rather than full data sets, a central question
is how to derive low-dimensional summary statistics from the observed
data with minimal loss of information. In this article we provide a
comprehensive review and comparison of the performance of the principal
methods of dimension reduction proposed in the ABC literature. The
methods are split into three nonmutually exclusive classes consisting
of best subset selection methods, projection techniques and
regularization. In addition, we introduce two new methods of dimension
reduction. The first is a best subset selection method based on Akaike
and Bayesian information criteria, and the second uses ridge regression
as a regularization procedure. We illustrate the performance of these
dimension reduction techniques through the analysis of three
challenging models and data sets.
\end{abstract}

% KEYWORDS
%

\end{frontmatter}

%s1 #&#
\section{Introduction}
%%%%%%%%%%%%%%%%%%%%%%%
%%%%%%%%%%%%%%%%%%%%%%

Bayesian inference is typically focused on the posterior distribution
$p(\theta|y_{\mathrm{obs}})\propto p(y_{\mathrm{obs}}|\theta)p(\theta)$ of a parameter
vector $\theta\in\Theta\subseteq\mathbb{R}^q$, $q\geq1$, representing
the updating of one's prior beliefs, $p(\theta)$, through the
likelihood (model) function, $p(y_{\mathrm{obs}}|\theta)$, having observed data
$y_{\mathrm{obs}}\in{\mathcal Y}$.
The term \textit{approximate Bayesian computation} (ABC) refers to a
family of models and algorithms that aim to draw samples from an
approximate posterior distribution when the likelihood,
$p(y_{\mathrm{obs}}|\theta)$, is unavailable or computationally intractable, but
where it is feasible to quickly generate data from the model, $y\sim
p(\cdot|\theta)$. ABC is rapidly becoming a popular tool for the
analysis of complex statistical models in an increasing number and
breadth of research areas. See, for example, \citet{lopes+b09},
\citet{bertorelle+bm10}, \citet{Beaumont10},
\citet{CsilleryEtAl10} and \citet{sisson+f11} for a partial
overview of the application of ABC methods.

ABC introduces two principal approximations to the posterior
distribution. First, the posterior distribution of the full data set,
$p(\theta|y_{\mathrm{obs}})$, is approximated by $p(\theta|s_{\mathrm{obs}})\propto
p(s_{\mathrm{obs}}|\theta)p(\theta)$, where $s_{\mathrm{obs}}=S(y_{\mathrm{obs}})$ is a vector of
summary statistics of lower dimension than the data $y_{\mathrm{obs}}$. In this
manner, $p(\theta|s_{\mathrm{obs}})\approx \break p(\theta|y_{\mathrm{obs}})$ is a good
approximation if $s_{\mathrm{obs}}$ is highly informative for the model
parameters, and $p(\theta|s_{\mathrm{obs}})=p(\theta|y_{\mathrm{obs}})$ if $s_{\mathrm{obs}}$ is
sufficient. As $p(s_{\mathrm{obs}}|\theta)$ is also likely to be computationally
intractable if $p(y_{\mathrm{obs}}|\theta)$ is computationally intractable, a
second approximation is constructed as $p_{\mathrm{ABC}}(\theta|s_{\mathrm{obs}})=\int
p(\theta,s|s_{\mathrm{obs}}) \,ds$, with
%
%e1 #&#
\begin{equation}
\label{eqnabcjointpost} p(\theta,s|s_{\mathrm{obs}}) \propto K_{\epsilon}\bigl(
\|s-s_{\mathrm{obs}}\|\bigr)p(s|\theta)p(\theta),
\end{equation}
where $K_\epsilon(\|u\|)=K(\|u\|/\epsilon)/\epsilon$ is a standard
smoothing kernel with scale parameter $\epsilon>0$.
As a result of (\ref{eqnabcjointpost}), approximating the target
$p(\theta|s_{\mathrm{obs}})$ by $p_{\mathrm{ABC}}(\theta|\allowbreak s_{\mathrm{obs}})$
can be shown to be a good approximation if the kernel scale parameter,
$\epsilon$, is small enough, following standard kernel density
estimation arguments (e.g., \cite{Blum10}).

In combination, both approximations allow for practical methods of
sampling from $p_{\mathrm{ABC}}(\theta|s_{\mathrm{obs}})$ that avoid explicit evaluation
of the intractable likelihood function, $p(y_{\mathrm{obs}}|\theta)$. A simple
rejection-sampling algorithm to achieve this was proposed by
\citet{PritchardEtAl99} (see also \cite{MarjoramEtAl03}),
which produces draws from $p(\theta,s|s_{\mathrm{obs}})$. In general terms, an
importance-sampling version of this algorithm proceeds as follows:
\begin{longlist}[(3)]
\item[(1)] Draw a candidate parameter vector from the prior, $\theta'\sim
p(\theta)$;
\item[(2)] Draw summary statistics from the model $s'\sim p(s|\theta')$;
\item[(3)] Assign to $(\theta',s')$ a weight, $w'$, that is proportional to
$K_{\epsilon}(\|s'-s_{\mathrm{obs}}\|)$.
\end{longlist}
Here, the sampling distribution for $(\theta',s')$ is the prior
predictive distribution, $p(s|\theta)p(\theta)$, and the target
distribution is $p(\theta,s|s_{\mathrm{obs}})$. Using equation (\ref
{eqnabcjointpost}), it is then straightforward to compute the
importance weight for the pair $(\theta',s')$. The weight is
proportional to $p(\theta',s'|s_{\mathrm{obs}})/[p(s'|\theta')p(\theta
')]=K_{\epsilon}(\|s'-s_{\mathrm{obs}}\|)$, which is free of intractable
likelihood terms, $p(s'|\theta')$. The manner by which the intractable
likelihoods cancel between sampling and target distributions forms the
basis for the majority of ABC algorithms.

Clearly, both ABC approximations to the posterior distribution help to
avoid the computational intractability of the original problem. The
first approximation allows the kernel weighting of the second
approximation, $K_\epsilon(\|s-s_{\mathrm{obs}}\|)$, to be performed on a lower
dimension than that of the original data, $y_{\mathrm{obs}}$. Kernel smoothing
is known to suffer from the curse of dimensionality (e.g.,
\cite{Blum10}), and so keeping $\dim(s)\leq\dim(y)$ as small as
possible helps to improve algorithmic efficiency. The second
approximation (\ref{eqnabcjointpost}) allows the sampler weights (or
acceptance probabilities, if one considers rejection-based samplers,
such as Markov chain Monte Carlo) to be free of intractable likelihood terms.

In practice, however, there is typically a trade-off between the two
approximations: if the dimension of $s$ is large so that the first
approximation, $p(\theta|s_{\mathrm{obs}})\approx p(\theta|y_{\mathrm{obs}})$, is good,
the second approximation may then be poor due to the inefficiency of
kernel smoothing in large dimensions.
Conversely, if the dimension of $s$ is small while the second
approximation (\ref{eqnabcjointpost}) will be good (with a small
kernel scale parameter, $\epsilon$), any loss of information in the
mapping $s_{\mathrm{obs}}=S(y_{\mathrm{obs}})$ means that the first approximation may be
poor. Naturally, a low-dimensional and near-sufficient statistic, $s$,
would provide a near-optimal and balanced choice.

For a given set of summary statistics, much work has been done on
deriving more efficient sampling algorithms to reduce the effect of the
second approximation by allowing a smaller value for the kernel scale
parameter, $\epsilon$, which in turn improves the approximation
$p_{\mathrm{ABC}}(\theta|s_{\mathrm{obs}})\approx p(\theta|s_{\mathrm{obs}})$. The greater the
algorithmic efficiency, the smaller the scale parameter that can be
achieved for a given computational burden. These algorithms include
Markov chain Monte Carlo (\cite{MarjoramEtAl03}; \cite{BortotEtAl07}) and
sequential Monte Carlo techniques (\cite{SissonEtAl07};
\cite{ToniEtAl09}; \cite{BeaumontEtAl09}; \cite{Drovandi+p11};
\cite{peters+fs12}; \cite{delmoral+dj12}).
By contrast, the regression-based methods described in Section~\ref
{sectionregression-adjustment} do not aim at reducing the scale
parameter $\epsilon$ but rather explicitly account for the imperfect
match between observed and simulated summary statistics
(\cite{BeaumontEtAl02}; \cite{BlumFrancois10}).

Achieving a good trade-off between the two approximations revolves
around the identification of a set of summary statistics, $s$, which
are both low-dimensional and highly informative for $\theta$. A number
of methods, primarily based on dimension reduction ideas, have been
proposed to achieve this (\cite{JoyceMarjoram08};
\cite{WegmannEtAl09}; \cite{NunesBalding10}; \cite{BlumFrancois10};
\cite{BlumCompstat10}; \cite{fernhead+p11}).
The choice of summary statistics is one of the most important aspects
of a statistical analysis using ABC methods (along with the choice of
algorithm). Poor specification of $s$ can have a large and detrimental
impact on both ABC model approximations.

In this article we provide the first detailed review and comparison of
the performance of the current methods of dimension reduction for
summary statistics within the ABC framework. We characterize these
methods into three nonmutually exclusive classes: (i) best subset
selection, (ii) projection techniques and (iii) regularization approaches.
As part of this analysis, we introduce two additional novel techniques
for dimension reduction within ABC. The first adopts the ideas of
Akaike and Bayesian information criteria to the ABC framework, whereas
the second makes use of ridge regression as a regularization procedure
for ABC.
The dimension reduction methods are compared through the analysis of
three challenging models and data sets. These involve the analysis of a
coalescent model with recombination (\cite{JoyceMarjoram08}), an
evaluation of the evolutionary fitness cost of mutation in
drug-resistant tuberculosis (\cite{LucianiEtAl09}) and an assessment
of the number and size-distribu\-tion of particle inclusions in the
production of clean steels (\cite{BortotEtAl07}).

The layout of this article is as follows: in Section~\ref
{sectionexisting} we classify and review the existing methods of
summary statistic dimension reduction in ABC, and in Section \ref
{sectionnew} we outline our two additional novel methods.
A comparative analysis of the performance of each of these methods is
provided in Section~\ref{sectionexamples}. We conclude with a discussion.

%%%%%%%%%%%%%%%%%%%%%%%%%%%%%%%%%%%%%%%%%%%%%%%%%%%%%%%%%%%
%%%%%%%%%%%%%%%%%%%%%%%%%%%%%%%%%%%%%%%%%%%%%%%%%%%%%%%%%%%%
%s2 #&#
\section{Classification of ABC Dimension Reduction Methods}
%%%%%%%%%%%%%%%%%%%%%%%%%%%%%%%%%%%%%%%%%%%%%%%%%%%%%%%%%%%%
%%%%%%%%%%%%%%%%%%%%%%%%%%%%%%%%%%%%%%%%%%%%%%%%%%%%%%%%%%%
\label{sectionexisting}

In a typical ABC analysis, an initial collection of statistics
$s^{\top}=(s_1,\ldots,s_p)$ is chosen by the modeler, the elements of
which have the potential to be informative for the model parameters,
$\theta^{\top}=(\theta_1,\ldots,\theta_q)$. Choice of these initial
statistics is highly problem specific, and the number of candidate
statistics, $p$, often considerably outnumbers the number of model
parameters, $q$, that is, $p\gg q$ (e.g.,
\cite{BortotEtAl07}; \cite{allingham+km09}; \cite{LucianiEtAl09}).
For example, \citet{BortotEtAl07} and \citet{allingham+km09}
use the ordered observations $S(y)=(s_{(1)},\ldots,s_{(p)})$ so that
there is no loss of information at this stage.
The analysis then proceeds by either using all $p$ statistics in full
or by attempting\vadjust{\goodbreak} to reduce their dimension while minimizing information
loss. Note that the most suitable set of summary statistics for an
analysis may be data set dependent, as the information content of
summary statistics may vary within the parameter space, $\Theta$ (an
exception is when sufficient statistics are known). As such, any
analysis should also consider establishing potentially different
summary statistics when re-implementing any model with a different data set.

Methods of summary statistics dimension reduction for ABC can be
broadly classified into three nonmutually exclusive classes.
The first class of methods follows a \textit{best subset selection}
approach. Here, candidate subsets are evaluated and ranked according to
various information-based criteria, such as measures of sufficiency
(\cite{JoyceMarjoram08}) or the entropy of the posterior distribution
(\cite{NunesBalding10}). In this article we contribute additional
criteria for this process derived from Akaike and Bayesian information
criteria arguments. From these criteria, the highest ranking subset
(or, alternatively, a subset consisting of those summary statistics
which demonstrate clear importance) is then chosen for the final analysis.

The second class of methods can be considered as \textit{projection
techniques}. Here, the
dimension of $(s_1,\ldots,\break s_p)$ is reduced by considering linear or
nonlinear combinations of the summary statistics. These methods make
use of a regression layer within the ABC framework, whereby the
response variable, $\theta$, is regressed by the (possibly transformed)
predictor variables, $s$ (\cite{BeaumontEtAl02}; \cite{BlumFrancois10}).
These projection methods include partial least squares regression
(Wegmann,\break Leuenberger and Excoffier (\citeyear{WegmannEtAl09})),
feed-forward neural networks (\cite
{BlumFrancois10}) and regression guided by minimum expected posterior
loss considerations (\cite{fernhead+p11}).

In this article we introduce a third class of methods for dimension
reduction in ABC, based on \textit{regularization techniques}. Using ridge
regression, we also make use of the regression layer between the
parameter $\theta$ and the summary statistics, $s$. However, rather
than explicitly considering a selection of summary statistics, we
propose to approach this implicitly, by shrinking the regression
coefficients toward zero so that uninformative summary statistics have
the weakest contribution in the regression equation.

In the remainder of this section we discuss each of these methods in
more detail. We first describe the ideas behind ABC regression
adjustment strategies (\cite{BeaumontEtAl02};\vadjust{\goodbreak} \cite{BlumFrancois10}), as
many of the dimension reduction techniques build on this framework.

%%%%%%%%%%%%%%%%%%%%%%%%%%%%%%%%%%%%%%%%%%
%%%%%%%%%%%%%%%%%%%%%%%%%%%%%%%%%%%%%%%%%%%
%s2.1 #&#
\subsection{Regression Adjustment in ABC}
%%%%%%%%%%%%%%%%%%%%%%%%%%%%%%%%%%%%%%%%%%%
%%%%%%%%%%%%%%%%%%%%%%%%%%%%%%%%%%%%%%%%%%
\label{sectionregression-adjustment}

Standard ABC methods suffer from the curse of dimensionality in that
the rate of convergence of posterior expectations with respect to
$p_{\mathrm{ABC}}(\theta|s_{\mathrm{obs}})$ (such as the Nadaraya--Watson estimator of the
posterior mean) decreases dramatically as the dimension of the summary
statistics, $p$, increases (\cite{Blum10}).
ABC regression adjustment (\cite{BeaumontEtAl02}) aims to avoid this
by explicitly modeling the discrepancy between $s$ and $s_{\mathrm{obs}}$. When
describing regression adjustment methods, for notational simplicity and
clarity of exposition, we assume that the parameter of interest,~$\theta$, is univariate (i.e., $q=1$). Regression adjustment methods may be
readily applied to multivariate $\theta$, by using a different
regression equation for each parameter, $\theta_1,\ldots,\theta_q$, separately.

The simplest model for this is a homoscedastic regression in the region
of $s_{\mathrm{obs}}$, so that
\[
\theta^i=m\bigl(s^i\bigr)+e^i,
\]
where $(\theta^i,s^i)\sim p(s|\theta)p(\theta)$ are $i=1,\ldots,n$
draws from the prior predictive distribution,
$m(s^i)=\mathbb{E}[\theta|s=s^i]$ is the mean function, and the $e^i$
are zero-mean random variates with common variance. To estimate the
conditional mean $m(\cdot)$, \citet{BeaumontEtAl02} assumed a
linear model
%
%e2 #&#
\begin{equation}
\label{eqlinear}
m\bigl(s^i\bigr)=\alpha+\beta^\top{s^i}
\end{equation}
in the neighborhood of $s_{\mathrm{obs}}$.
An estimate of the mean function, $\hat{m}(\cdot)$, is obtained by
minimizing the weight\-ed least squares criterion $\sum_{i=1}^n w^i\|
m(s^i)-\theta^i \|^2$,\break where $w^i=K_\epsilon(\|s^i-s_{\mathrm{obs}}\|)$. A
weight\-ed sample from the posterior distribution, $p_{\mathrm{ABC}}(\theta
|s_{\mathrm{obs}})$, is then obtained by the adjustment
%
%e3 #&#
\begin{equation}
\label{eqcorrec} \theta^{*i}=\hat{m}(s_{\mathrm{obs}})+\bigl(
\theta^i-\hat{m}\bigl(s^i\bigr)\bigr)
\end{equation}
for $i=1,\ldots,n$. In the above, the kernel scale parameter $\epsilon$
controls the bias-variance trade-off: increasing $\epsilon$ reduces
variance by increasing the effective sample size---the number of
accepted simulations when using a uniform kernel $K$---but increases
bias arising from departures from a linear mean function $m(\cdot)$ and
homoscedastic error structure (\cite{Blum10}).

\citet{BlumFrancois10} proposed the more flexible, heteroscedastic model
%
%e4 #&#
\begin{equation}
\label{eqhetero2} \theta^i=m\bigl(s^i\bigr)+\sigma
\bigl(s^i\bigr)e^i,
\end{equation}
where $\sigma^2(s^i)=\mathbb{V}[\theta|s=s^i]$ denotes the conditional
variance. This\vadjust{\goodbreak} variance is estimated using a second regression model
for the log of the squared residuals, that is, $\log(\theta^i-\hat
{m}(s^i))^2=\log\sigma^2(s^i) + \eta^i$, where the $\eta^i$ are
independent, zero-mean variates with common variance. The equivalent
adjustment to (\ref{eqcorrec}) is then given by
%
%e5 #&#
\begin{equation}
\label{eqhetero} \theta^{*i}=\hat{m}(s_{\mathrm{obs}})+ \bigl[
\theta^i-\hat{m}\bigl(s^i\bigr) \bigr]\frac{\hat{\sigma}(s_{\mathrm{obs}})}{\hat
{\sigma}(s^i)},
\end{equation}
where $\hat{\sigma}(s)$ denotes the estimate of $\sigma(s)$. The kernel
scale parameter, $\epsilon$, plays the same role as for the
homoscedastic model, except with more flexibility on deviations from
homoscedasticity.
\citet{nott+fms11} have demonstrated that regression adjustment
ABC algorithms produce samples, $\{\theta^{*i}\}$, for which first- and
second-order moment summaries approximate adjusted expectation and
variance for a Bayes linear analysis. We do not describe here an
alternative regression adjustment method where the summary statistics
are rather considered as the dependent variables and the parameters as
the independent variables of the regression
(\cite{Leuenberger+w10}).

%%%%%%%%%%%%%%%%%%%%%%%%%%%%%%%%%%%%%%%%%%%
%%%%%%%%%%%%%%%%%%%%%%%%%%%%%%%%%%%%%%%%%%%%
%s2.2 #&#
\subsection{Best Subset Selection Methods}
%%%%%%%%%%%%%%%%%%%%%%%%%%%%%%%%%%%%%%%%%%%%
%%%%%%%%%%%%%%%%%%%%%%%%%%%%%%%%%%%%%%%%%%%

Best subset selection methods are conceptually simple, but are
cumbersome to manage for large numbers of potential summary statistics,
$s=(s_1,\ldots,\break s_p)$. Exhaustive enumeration of the $2^p-1$ possible
combinations of summary statistics is practically infeasible beyond a
moderate value of $p$. This is especially true of Markov chain Monte
Carlo or sequential Monte Carlo based analyses, which require one
sampler implementation per combination. As a result, stochastic or
deterministic (greedy) search procedures, such as forward or backward
selection, are required to implement them.

%%%%%%%%%%%%%%%%%%%%%%%%%%%%%%%%%%%%%%%%
%%%%%%%%%%%%%%%%%%%%%%%%%%%%%%%%%%%%%%%%%
%s2.2.1 #&#
\subsubsection{A sufficiency criterion}
%%%%%%%%%%%%%%%%%%%%%%%%%%%%%%%%%%%%%%%%%
%%%%%%%%%%%%%%%%%%%%%%%%%%%%%%%%%%%%%%%%
\label{sectionJoyceMarjoram}

The first principled approach to dimension reduction in ABC was the
$\varepsilon$-sufficiency concept proposed by \citet{JoyceMarjoram08},
which was used to determine whether to include an additional summary
statistic, $s_k$, to a model already containing statistics
$s_1,\ldots,s_{k-1}$. Here, noting that the difference between the log
likelihoods of $p(s_1,\ldots,s_k|\theta)$ and
$p(s_1,\ldots,s_{k-1}|\theta)$ is\break $\log
p(s_k|s_1,\ldots,s_{k-1},\theta)$, \citet{JoyceMarjoram08} defined the
set of statistics $s_1,\ldots,s_{k-1}$ to be $\varepsilon$-suffi\-cient
relative to $s_k$ if
%
%e6 #&#
\begin{eqnarray}
\label{eqne-sufficient}
\delta_k &=& \sup_\theta\log
p(s_k|s_1,\ldots,s_{k-1},\theta) \nonumber\\
&&{}-
\inf_\theta\log p(s_k|s_1,\ldots,s_{k-1},
\theta) \\
&\leq&\varepsilon.\nonumber
\end{eqnarray}
Accordingly, if an estimate of $\delta_k$ (i.e., the ``score'' of $s_k$
relative to $s_1,\ldots,s_{k-1}$) is greater than $\varepsilon$, then
there is enough additional information content in $s_k$ to justify
including it in the model. In practice, \citet{JoyceMarjoram08}
implement a conceptually equivalent assessment, whereby $s_k$ is added
to the model if the ratio of posteriors
\[
R_k(\theta) = \frac{p_{\mathrm{ABC}}(\theta| s_1,\ldots
,s_{k-1},s_k)}{p_{\mathrm{ABC}}(\theta| s_1,\ldots, s_{k-1})} %\label{eqR}
\]
differs from one by more than some threshold value $T(\theta)$ for any
value of $\theta$. As such, a statistic $s_k$ will be added to the
model if the resulting posterior changes sufficiently at any point.
The threshold, $T(\theta)$, is user-specified, with one particular choice
described in Section 5 of \citet{JoyceMarjoram08}.

This procedure can be implemented within any stepwise search algorithm,
each of which have various pros and cons.
Following the definition (\ref{eqne-sufficient}), the resulting
optimal subset of summary statistics is then $\varepsilon$-sufficient
relative to each one of the remaining summary statistics.
Here $\varepsilon$ intuitively represents an acceptable error in
determining whether $s_k$ contains further useful information in
addition to $s_1,\ldots,\break s_k$. This quantity is also user-specified, and
so the final optimal choice of summary statistics will depend on the
chosen value.

Sensitivity to the choice of $\varepsilon$ aside, this approach may be
criticized in that it assumes that every change to the posterior
obtained by adding a statistic, $s_k$, is beneficial. It is conceivable
that attempting to include a completely noninformative statistic, where
the observed statistic is unlikely to have been generated under the
model, will result in a sufficiently modified posterior as measured by
$\varepsilon$, but one which is more biased away from the true
posterior $p(\theta|y_{\mathrm{obs}})$ than without including $s_k$. A toy
example illustrating this was given by \citet{sisson+f11}.

A further criticism is that the amount of computation required to
evaluate $R_k(\theta)$ for all $\theta$, and on multiple occasions, is
considerable, especially for large $q$. In practice,
\citet{JoyceMarjoram08} considered $\theta$ to be univariate, and
approximated continuous $\theta$ over a discrete grid in order to keep
computational overheads to acceptable levels. As such, this method
appears largely restricted to dimension reduction for univariate
parameters ($q=1$).

%%%%%%%%%%%%%%%%%%%%%%%%%%%%%%%%%%%%%
%%%%%%%%%%%%%%%%%%%%%%%%%%%%%%%%%%%%%%
%s2.2.2 #&#
\subsubsection{An entropy criterion}
%%%%%%%%%%%%%%%%%%%%%%%%%%%%%%%%%%%%%%
%%%%%%%%%%%%%%%%%%%%%%%%%%%%%%%%%%%%%
\label{secentropy}

Nunes and Balding\break (\citeyear{NunesBalding10}) propose the entropy of a
distribution as a heuristic to measure\vadjust{\goodbreak} the informativeness of candidate
combinations of summary statistics.
Since entropy measures information and a lack of randomness (\cite
{shannon+w48}), the authors propose minimizing the entropy of the
approximate posterior, $p_{\mathrm{ABC}}(\theta|s_{\mathrm{obs}})$, over subsets of the
summary statistics, $s$, as a proxy for determining maximal information
about a parameter of interest. High entropy results from a diffuse
posterior sample, whereas low entropy is obtained from a posterior
which is more precise in nature.

\citet{NunesBalding10} estimate entropy using the unbiased $k$th
nearest neighbor estimator of \citet{SinghEtAl03}. For a weighted
posterior sample, $(w^1,\theta^1),\ldots,(w^n,\theta^n)$, where
$\sum_iw^i=1$, this estimator can be written as
%
%e7 #&#
\begin{eqnarray}
\label{entent} %\hat{E}=\log\left[\frac{\pi^{q/2}}{\Gamma(q/2{+}1)}
\hat{E}&=&\log\biggl[\frac{\pi^{q/2}}{\Gamma(q/2{+}1)} \biggr] -
\psi(k) +\log n\nonumber\\[-8pt]\\[-8pt]
&&{}+q \sum_{i=1}^{n}
w^{i}\log\hat{C}_i^{-1}\bigl(k/(n-1)\bigr),\nonumber
\end{eqnarray}
where\vspace*{1pt} $q=\dim(\theta)$, $\psi(x)=\Gamma'(x)/\Gamma(x)$ denotes the
digamma function, and where $\hat{C}_i(\cdot)$ denotes the empirical
distribution function of the Euclidean distance from $\theta^i$ to the
remainder of the weighted posterior sample, that is, of the weighted
samples $\{(\tilde{w}^j,\break \|\theta^i-\theta^j\|)\}_{j\neq i}$, where
$\tilde{w}^j=w^j/\sum_{j\neq i}w^j$.
Following \citet{SinghEtAl03}, the original work of
\citet{NunesBalding10} used $k=4$ and was based on an equally weighted
posterior sample (i.e., with $w^i=1/n, i=1,\ldots,n$), so that
$\hat{C}_i^{-1}(k/(n-1))$ denotes the Euclidean distance from
$\theta^i$ to its $k$th closest neighbor in the posterior sample
$\{\theta^1,\ldots,\theta^{i-1},\theta^{i+1},\ldots,\theta^n\}$.

While minimum entropy could in itself be used to evaluate the
informativeness of a vector of summary statistics for $\theta$
(although see the criticism of entropy below),
\citet{NunesBalding10} propose a second stage to their analysis,
which aims to assess the performance of a candidate set of summary
statistics using a measure of posterior error. For example, when the
true parameter vector, $\theta_{\mathrm{true}}$, is known, the authors suggest
the root sum of squared errors (RSSE), given by
%
%e8 #&#
\begin{equation}
\label{eqnrsse} \mbox{RSSE} = \Biggl(\sum_{i=1}^n
w^i\bigl\|\theta^i-\theta_{\mathrm{true}}\bigr\|^2
\Biggr)^{1/2},
\end{equation}
where the measure $\|\theta^i-\theta_{\mathrm{true}}\|$ compares the components
of $\theta$ on a suitable scale (and so some component-wise
standardization may be required). Naturally, the true parameter value,
$\theta_{\mathrm{true}}$, is unknown in practice. However, if the simulated
summary statistics from the samples $(\theta^i,s^i)$ are treated as
observed data, it is clear that $\theta_{\mathrm{true}}=\theta^i$ for the
posterior\break $p_{\mathrm{ABC}}(\theta|s^i)$. As such, the RSSE can be easily
computed with a leave-one-out technique.

As the subset of summary statistics that minimizes (\ref{eqnrsse})
will likely vary over observed data sets, $s^i$,
\citet{NunesBalding10} propose minimizing the average RSSE over
some number of simulated data sets which are close to the observed,
$s_{\mathrm{obs}}$. To avoid circularity, \citet{NunesBalding10} define
these ``close'' data sets to be the $j=1,\ldots,n^*$ simulated data
sets, $\{s^j\}$, that minimize $\|s^j_{\mathrm{ME}}-s_{\mathrm{ME}}\|$, where $s^j_{\mathrm{ME}}$
and $s_{\mathrm{ME}}$ are the vectors of minimum entropy summary statistics
computed via (\ref{entent}) from $s^j$ and the observed summary
statistics, $s_{\mathrm{obs}}$, respectively. That is, the quantity
%
%e9 #&#
\begin{equation}
\label{eqnmrsse} \overline{\mbox{RSSE}}=\frac{1}{n^*}\sum
_{j=1}^{n^*} \mbox{RSSE}_j
\end{equation}
is minimized (over subsets of summary statistics), where
$\mbox{RSSE}_j$ corresponds to (\ref{eqnrsse}) using the simulated
data set $s^j$.

This approach is intuitive and is attractive because the second stage
directly measures error in the posterior with respect to a known truth,
$\theta_{\mathrm{true}}$, which is not typically considered in other ABC
dimension reduction approaches, albeit at the extra computational
expense of a two-stage procedure. A weakness of the first stage,
however, is the assumption that addition of an informative statistic
will reduce the entropy of the resulting posterior distribution. An
example of when this does not occur is when the posterior distribution
is diffuse with respect to the prior---for instance, if an overly
precise prior is located in the distributional tails of the posterior
(e.g., \cite{jeremiah+smms11}). In this case, attempting to
include an informative additional statistic, $s_k$, can result in a
distribution that is more diffuse than with $s_k$ excluded. As such,
the entropic approach is therefore mostly suited to models with
relatively diffuse prior distributions.
Another potential criticism of the first stage is that minimizing the
entropy does not necessarily provide the \textit{minimal} subset of
sufficient statistics. This provides an argument for considering the
mutual information between $\theta$ and~$s$, rather than the entropy
(\cite{barnesetal11}; see also \cite{Filippi12}).
However, it is clear that the overall approach of
\citet{NunesBalding10} could easily be implemented with alternative
first-stage selection criteria.

%%%%%%%%%%%%%%%%%%%%%%%%%%%%%%%%%%%%%
%%%%%%%%%%%%%%%%%%%%%%%%%%%%%%%%%%%%%%
%s2.2.3 #&#
\subsubsection{AIC and BIC criteria}
%%%%%%%%%%%%%%%%%%%%%%%%%%%%%%%%%%%%%%
%%%%%%%%%%%%%%%%%%%%%%%%%%%%%%%%%%%%%

Information criteria based on Akaike and Bayesian information are
natural best subset selection techniques for summary statistic
dimension reduction in ABC analyses. We introduce and develop these
criteria in Section~\ref{sectionnewaicbic}.

%%%%%%%%%%%%%%%%%%%%%%%%%%%%%%%%%%
%%%%%%%%%%%%%%%%%%%%%%%%%%%%%%%%%%%
%s2.3 #&#
\subsection{Projection Techniques}
%%%%%%%%%%%%%%%%%%%%%%%%%%%%%%%%%%%
%%%%%%%%%%%%%%%%%%%%%%%%%%%%%%%%%%

Selecting a best subset of summary statistics from $s=(s_1,\ldots,s_p)$
suffers from the problem that it may require several statistics to
provide the same information content as a single, highly informative
statistic that was not specified in the initial set, $s$. To avoid
this, projection techniques aim to combine the elements of $s$ through
linear or nonlinear transformations, in order to construct a
potentially much lower-dimensional set of highly informative statistics.

One of the main advantages of projection techniques is that, unlike
best subset selection methods, they scale well with increasing numbers
of summary statistics. They can handle large numbers of possibly
uninformative summary statistics, in addition to accounting for high
levels of interdependence and multicollinearity. A~minor disadvantage
of projection techniques is that the final sets of projected summary
statistics typically (but not universally) lack interpretability. In
addition, most projection methods require the specification of a
hyperparameter that governs the number of projections to perform.

%%%%%%%%%%%%%%%%%%%%%%%%%%%%%%%%%%%%%%%%%%%%%%%%
%%%%%%%%%%%%%%%%%%%%%%%%%%%%%%%%%%%%%%%%%%%%%%%%%
%s2.3.1 #&#
\subsubsection{Partial least squares regression}
%%%%%%%%%%%%%%%%%%%%%%%%%%%%%%%%%%%%%%%%%%%%%%%%%
%%%%%%%%%%%%%%%%%%%%%%%%%%%%%%%%%%%%%%%%%%%%%%%%

Partial least squares regression seeks the orthogonal linear
combinations of the explanatory variables which have high variance and
high correlation with the
response variable (e.g., \cite{boulesteix+s07}; \cite{vinzi+chw10};
\cite{AbdiWilliams10}). \citet{WegmannEtAl09} proposed the use of
partial least squares regression for dimension reduction in ABC, where
the explanatory variables are the suitably (e.g., Box--Cox) transformed
summary statistics, $s$, and the response variables is the parameter
vector, $\theta$.

The output of a partial least squares analysis is the set of $k$
orthogonal components of the regression design matrix
%
%e10 #&#
\begin{equation}
X=\label{eqdesign} %\[
\pmatrix{ 1 & s^1_1&
\cdots& s_p^1
\cr
\vdots& \vdots& \ddots& \vdots
\cr
1
& s^n_1 & \cdots& s^n_p}
\end{equation}
that are optimally correlated (in a specific sense) with~$\theta$.
Here, $s_j^i$ denotes the $j$th component of the $i$th simulated
summary statistic, $s^i$.\vadjust{\goodbreak}
To choose the appropriate number of orthogonal components, \citet
{WegmannEtAl09}
examine the root mean square error of $\theta$ for each value of~$k$,
as estimated by a leave-one-out cross-validation strategy. For a fixed
number of components, $k$, this corresponds to
%
%e11 #&#
\begin{equation}
\label{eqnPLSLOO} \mbox{RMSE}_{k}= \Biggl(\frac{1}{n}\sum
_{i=1}^{n} \bigl\|\hat{m}^{-i}_k
\bigl(s^i\bigr)-\theta^i\bigr\|^2
\Biggr)^{1/2},
\end{equation}
where $\hat{m}^{-i}_k(s)$ denotes the mean response of the partial
least squares regression, estimated without the $i$th simulated summary
statistic, $s^i$ (e.g., \cite{mevik+c04}). The optimal number of
components is then chosen by inspection of the $\mbox{RMSE}_k$ values,
based on minimum gradient change arguments (e.g., \cite{mevik+w07}).

A potential disadvantage of partial least squares regression, as
performed by \citet{WegmannEtAl09}, is that it aims to infer a
global linear relationship between $\theta$ and $s$ based on draws from
the prior predictive distribution, $p(s|\theta)p(\theta)$. This may
differ from the relationship observed in the region around $s=s_{\mathrm{obs}}$,
and as such may produce unsuitable orthogonal components as a result. A
workaround for this would be to follow \citet{fernhead+p11} (see
Section~\ref{sectionFP11}) and elicit the relationship between
$\theta$ and $s$ based on samples from a truncated prior
$(\theta^i,s^i)\sim p(s|\theta)p(\theta)I(\theta\in\Theta^R)$, where
$\Theta^R\subset\Theta$ restricts the samples, $\theta^i$, to regions
of significant posterior density. One simple way to identify such a
region is through a pilot ABC analysis (\cite{fernhead+p11}).

%%%%%%%%%%%%%%%%%%%%%%%%%%%%%%%
%%%%%%%%%%%%%%%%%%%%%%%%%%%%%%%%
%s2.3.2 #&#
\subsubsection{Neural networks}
%%%%%%%%%%%%%%%%%%%%%%%%%%%%%%%%
%%%%%%%%%%%%%%%%%%%%%%%%%%%%%%%
\label{sectionneural-networks}

In the regression setting, feed-forward neural networks can be
considered as a nonlinear generalization of the partial least squares
regression technique described above. Blum and
Fran\-{\c{c}}ois (\citeyear{BlumFrancois10})
proposed the neural network as a machine learning approach to dimension
reduction by estimating the conditional mean and variance functions,
$m(\cdot)$ and $\sigma^2(\cdot)$ in the nonlinear, heteroscedastic
regression adjustment model (\ref{eqhetero2})---see Section
\ref{sectionregression-adjustment}.

The neural network reduces the dimension of the summary statistics to
$H<p$, using $H$ hidden units in the network, $z_1,\ldots,z_H$, defined as
%
%e12 #&#
\begin{equation}
\label{eqnffnn1} z_j=h \Biggl(\sum_{k=1}^p
\omega_{jk}^{(1)} s_k + \omega_{j0}^{(1)}
\Biggr)
\end{equation}
for $j=1,\ldots,H$.
The $\omega_{jk}^{(1)}$ terms are the weights of the first layer of the
neural network, and $h(\cdot)$ is a nonlinear function, typically\vadjust{\goodbreak} the
logistic function. The reduced and nonlinearly transformed summary
statistics of the hidden units, $z_j$, are then combined through the
regression function of the neural network
%
%e13 #&#
\begin{equation}
\label{eqnffnn2} m(s)=g \Biggl(\sum_{j=1}^H
\omega_{j}^{(2)} z_j +\omega_{0}^{(2)}
\Biggr),
\end{equation}
where $\omega_{j}^{(2)}$ denotes the weights of the second layer of the
neural network and $g(\cdot)$ is a link function. A similar neural
network is used to model $\log\sigma^2(s)$ (e.g., \cite{nix+w95}),
with the possibility of allowing for a different number of hidden units
to estimate heteroscedasticity in the regression adjustment compared to
that in the mean function $m(s)$.

Rather than dynamically determining the number of hidden units $H$,
\citet{BlumFrancois10} propose to specify a fixed value, such as
$H=q$ where $q=\dim(\theta)$ is the number of parameters to infer. The
weights of the neural network are then obtained by minimizing the
regularized least-squares criterion
\[
\sum_{i=1}^n w^i\bigl\|m
\bigl(s^i\bigr)-\theta^i \bigr\|^2 + \lambda\|
\omega\|^2,
\]
where $\omega$ is the vector of all weights in the neural network model
for $m(s)$, $w^i=K_\epsilon(\|s^i-s_{\mathrm{obs}}\|)$ is the weight of the
sample $(\theta^i,s^i)\sim p(s|\theta)p(\theta)$, and $\lambda>0$
denotes the regularization parameter (termed the weight-decay parameter
for neural networks). The idea of regularization is to shrink the
weights toward zero so that only informative summary statistics
contribute in the models (\ref{eqnffnn1}) and (\ref{eqnffnn2}) for
$m(s)$. Following the estimation of $m(s)$, a similar regularization
criterion is used to estimate $\log\sigma^2(s)$. Both mean and variance
functions can then be used in the regression adjustment of equation
(\ref{eqhetero}).

%%%%%%%%%%%%%%%%%%%%%%%%%%%%%%%%%%%%%%%%%%%%%%%
%%%%%%%%%%%%%%%%%%%%%%%%%%%%%%%%%%%%%%%%%%%%%%%%
%s2.3.3 #&#
\subsubsection{Minimum expected posterior loss}
%%%%%%%%%%%%%%%%%%%%%%%%%%%%%%%%%%%%%%%%%%%%%%%%
%%%%%%%%%%%%%%%%%%%%%%%%%%%%%%%%%%%%%%%%%%%%%%%
\label{sectionFP11}

Fearnhead and Prangle (\citeyear{fernhead+p11}) proposed a
decision-theoretic dimension reduction method with a slightly different
aim to previous dimension reduction approaches. Here, rather than
constructing appropriate summary statistics to ensure that
$p_{\mathrm{ABC}}(\theta |s_{\mathrm{obs}})\approx
p(\theta|y_{\mathrm{obs}})$ is a good approximation,
$p_{\mathrm{ABC}}(\theta|s_{\mathrm{obs}})$ is alternatively required
to be a good approximation in terms of the accuracy of specified
functions of the model parameters. In particular, assuming that
interest is in point estimates of the model parameters, if
$\theta_{\mathrm{true}}$ denotes the true parameter value and $\hat{\theta}$ an
estimate, then \citet {fernhead+p11} propose to choose those
summary statistics that minimize the quadratic loss
\[
L(\theta_{\mathrm{true}},\hat{\theta}) = (\theta_{\mathrm{true}}-\hat{
\theta})^\top A(\theta_{\mathrm{true}}-\hat{\theta})\vadjust{\goodbreak}
\]
for some $p\times p$ positive-definite matrix $A$. This loss is
minimized for $s_{\mathrm{obs}}=E_{p(\theta|y_{\mathrm{obs}})}(\theta)$, the true
posterior mean.

To estimate $E_{p(\theta|y)}(\theta)$, Fearnhead and Prangle\break
(\citeyear{fernhead+p11})
propose least squares regression models for the $k=1,\ldots,q$ model
parameters, $(\theta_1,\ldots,\theta_q)$, given by
%
%e14 #&#
\begin{equation}
\label{fp11}\qquad \theta_k^i=E_{p(\theta|y)}(
\theta_k)+\eta_k^i = \alpha_k +
\beta_k^\top f\bigl(s^i\bigr)+
\eta_k^i,
\end{equation}
where $(\theta^i,s^i)\sim p(s|\theta)p(\theta)$ are draws from the
prior predictive distribution, $\alpha_k$ and $\beta_k$ are unknown
regression parameters to be estimated, and $\eta_k^i$ denotes a
zero-mean noise process.
Here $f(s)$ is a vector of potentially nonlinear transformations of the
data (i.e., of the original summary statistics). For example, in one
application, \citet{fernhead+p11} use the polynomial basis
functions $f(s)=(s, s^2, s^3, s^4)$, that is, a~vector of length $4p$,
where $p=\dim(s)$ is the number of elements in $s$, consisting of the
first four powers of each element of $s$. The choice of $f(s)$ can be
based on standard diagnostics of regression fit, such as BIC. If the
prior $\pi(\theta)$ is diffuse with respect to the posterior, then one
may estimate the regression model (\ref{fp11}) based on samples from a
truncated prior $(\theta^i,s^i)\sim p(s|\theta)p(\theta)I(\theta\in
\Theta^R)$, where $\Theta^R\subset\Theta$ restricts the samples, $\theta
^i$, to regions of significant posterior density (e.g., via a pilot ABC
analysis). Clearly, more sophisticated alternatives to least squares
regression may be used.

After fitting equation (\ref{fp11}), the new, single summary statistic
for the parameter $\theta_k$ is $\hat{\beta}_k^\top f(s)$, where
$\hat{\beta}_k$ denotes the least squares estimate of $\beta_k$. The
resulting $q$-dimensional vector of new summary statistics is then used
in a standard ABC analysis. \citet{fernhead+p11} show that these new
statistics can lead to posterior inferences that considerably
outperform inferences based on the original statistics, $s$.
\citet{nott+fs12} demonstrate that these summary statistics can be
viewed as Bayes linear estimates of the posterior mean.

%%%%%%%%%%%%%%%%%%%%%%%%%%%%%%%%%%%%%%
%%%%%%%%%%%%%%%%%%%%%%%%%%%%%%%%%%%%%%%
%s2.4 #&#
\subsection{Regularization Approaches}
%%%%%%%%%%%%%%%%%%%%%%%%%%%%%%%%%%%%%%%
%%%%%%%%%%%%%%%%%%%%%%%%%%%%%%%%%%%%%%

Regularization approaches aim to reduce overfitting in a model by
penalizing model complexity. A~simple example where overfitting can
occur in ABC is the standard regression adjustment
(\cite{BeaumontEtAl02}; Section
\ref{sectionregression-adjustment}), where there is a risk of over
adjusting the parameters, $\theta^i$, in the direction of
uninformative summary statistics via (\ref{eqcorrec}). Regularization
is used as part of the estimation of the neural network weights in the
projection technique proposed by \citet{BlumFrancois10} (see Section
\ref{sectionneural-networks}). As such, the regression adjustment of
\citet{BeaumontEtAl02} is a procedure that could greatly benefit
from the inclusion of regularization techniques. We introduce the ridge
regression adjustment to ABC in Section~\ref{sectionridge}.

%%%%%%%%%%%%%%%%%%%%%%%%%%
%%%%%%%%%%%%%%%%%%%%%%%%%%%
%s2.5 #&#
\subsection{Other Methods}
%%%%%%%%%%%%%%%%%%%%%%%%%%%
%%%%%%%%%%%%%%%%%%%%%%%%%%

There are a number of alternative approaches to dimension reduction for
ABC, including methods that aim to circumvent the dimensionality issue,
that we do not include in our comparative analysis
(Section~\ref{sectionexamples}). \citet{drovandi+pf11} proposed to
adopt ideas from indirect inference (e.g., \cite{heggland+f04}) as a
means to identify summary statistics for an ABC analysis. This involves
specification of a model $\tilde{p}(\cdot|\tilde{\theta})$ which is
similar to $p(\cdot|\theta)$, but which is computationally tractable.
The idea is that estimates of $\tilde{\theta}$ under
$\tilde{p}(\cdot|\tilde{\theta})$, such as maximum likelihood estimates
or posterior means, are likely to be informative about $\theta$ if
$p(\cdot|\theta)$ and $\tilde{p}(\cdot|\tilde{\theta})$ are
sufficiently similar. This approach can be considered similar in spirit
to that of \citet{fernhead+p11} which uses estimated posterior
means under a pilot ABC analysis (see Section~\ref{sectionFP11}).
\citet{BlumCompstat10} proposed a Bayesian criterion related to
the BIC (see Section~\ref{sectionnewaicbic}) as a best subset selection
procedure. The idea is to implement a Bayesian analysis of the standard
regression adjustment mod\-el~(\ref{eqcorrec}). The criterion, called the
\textit{evidence} approximation, seeks the best subset of summary
statistics to regress the parameter $\theta$. In comparison to the BIC,
the evidence criterion is attractive because it contains no
approximation in its derivation. However, the downside is that its
computation requires the tuning of the Bayesian linear regression
hyperparameters.
Additionally,
\citet{AesBeaFut} proposed to use boosting for choosing summary statistics and
\citet{Jung2011} developed a genetic algorithm that\break
weights the summary statistics so that individual statistics do not
contribute equally to the comparisons between observations and
simulations. The aim is that the uninformative summary statistics
should ideally have negligible weights.

Finally, a number of recent ABC modeling approaches have attempted to
find ways of accurately handling the full vector of initial statistics,
$s$ [or~the full data set, $s=S(y)=y$], thereby avoiding the need to
perform dimension reduction. \citet{bonassi11} propose\vadjust{\goodbreak} fitting a
$(p+q)$-dimen\-sional mixture of Gaussian distributions to the sample
$(\theta^i,s^i)\sim p(s|\theta)p(\theta)$, $i=1,\ldots,n$, and then
finding the distribution of $\theta|s_{\mathrm{obs}}$ by conditioning on
observing $s=s_{\mathrm{obs}}$. This approach potentially requires a large
number of mixture components to accurately model the joint density when
$(p+q)$ is large. \citet{fan+ns12} suggest using an approximation
to $p(s|\theta)$ by approximating each marginal likelihood function,
$p(s_i|\theta)$, using a mixture of experts model, where the weights,
mean and variance of each mixture component is allowed to depend on
$\theta$, and then inducing dependence between these marginals using a
mixture of multivariate Gaussian distributions. This approach requires
continuous summary statistics for the mixture regression and is
practically useful for moderate $p$ (i.e., hundreds of summary
statistics). Writing $y=(y_1,\ldots,y_p)$, \citet{barthelme+c11}
propose to factorize the likelihood as $p(y|\theta)=\prod_i
p(y_i|y_{1:i-1},\theta)$ and construct an ABC approximation of each
component in turn [i.e., $p_{\mathrm{ABC}}(y_i|y_{1:i-1})\!=\!\int
K_\epsilon(\|y_i-y_{\mathrm{obs},i}\|)p(y_i|y_{1:i-1},\theta)\,dy_i$] with
computation performed using an expectation-propagation algorithm
(\cite{minka01}). This ap-\break proach,~while potentially fast and accurate,
assumes that conditional simulation of $y_i\sim
p(y_i|y_{1:i-1},\theta)$ is available for $i=1,\ldots,n$, and so is not
suitable for all models and analyses. Last, \citet{Jasra12}
exploit the structure of hidden Markov models to perform an iterative
sequence of ABC analyses, each using only a single data point in each
analysis, and \citet{Nakagome12} propose a novel approach to
post-processing ABC importance sampling output whose convergence rate
is claimed to avoid the curse of dimensionality.

%%%%%%%%%%%%%%%%%%%%%%%%%%%%%%%%%%%%%%%%%
%%%%%%%%%%%%%%%%%%%%%%%%%%%%%%%%%%%%%%%%%%
%s3 #&#
\section{New Dimension Reduction Methods}
%%%%%%%%%%%%%%%%%%%%%%%%%%%%%%%%%%%%%%%%%%
%%%%%%%%%%%%%%%%%%%%%%%%%%%%%%%%%%%%%%%%%
\label{sectionnew}

In this section we introduce two new dimension reduction criteria for
ABC methods. The first is a best subset selection procedure deriving
from AIC and BIC criteria, constructed under implementation of the
local linear model of equation (\ref{eqlinear})
(\cite{BeaumontEtAl02}). A similar idea was proposed and tested for
a Gaussian model by \citet{Sedki12}. The second is a modification
to the fitting of (\ref{eqlinear}) by considering ridge regression
instead of least squares regression. Both of these methods are now
implemented in the freely available \texttt{R} package \texttt{abc}
(\cite{csillery+fb11}).

%%%%%%%%%%%%%%%%%%%%%%%%%%%%%%%%%
%%%%%%%%%%%%%%%%%%%%%%%%%%%%%%%%%%
%s3.1 #&#
\subsection{AIC and BIC Criteria}
%%%%%%%%%%%%%%%%%%%%%%%%%%%%%%%%%%
%%%%%%%%%%%%%%%%%%%%%%%%%%%%%%%%%
\label{sectionnewaicbic}

Akaike information criterion (AIC) and Bayesian information criterion
(BIC) provide a measure of the relative goodness of fit of a
statistical model. Each can be expressed as the sum of the maximized
log-likelihood that measures the fit of the model to the data, and a
penalty for model complexity (\cite{Akaike74}; \cite{Schwarz78}). While
evaluation of $\log p(y_{\mathrm{obs}}|\hat{\theta}_{\mathrm{mle}})$ or $\log
p(s_{\mathrm{obs}}|\hat{\theta}_{\mathrm{mle}})$ is unavailable in the ABC framework and
determination of the maximum likelihood estimator, $\hat{\theta
}_{\mathrm{mle}}$, is challenging, a simple and tractable likelihood function is
available though the local-linear regression model of equation~(\ref
{eqlinear}) (Section~\ref{sectionregression-adjustment}).

Specifically, we consider the local linear regression model equation
(\ref{eqlinear}) of \citet{BeaumontEtAl02} for each parameter
$\theta_1,\ldots,\theta_q$ and assume independent Gaussian errors,
$e^j\sim N(0,\sigma_j^2)$, for $j=1,\ldots,q$. Then the AIC becomes
%
%e15 #&#
\begin{equation}
\label{eqAIC} \mbox{AIC}=\tilde{n}\log{\prod_{j=1}^q
\hat{\sigma}_j^2} +2d,
\end{equation}
where $d=q(p+1)$ is the number of estimated regression parameters and
$\tilde{n}$ is the \textit{effective} number of simulations used in the
local-linear regression model, which we define as $\tilde{n}=\sum
_{i=1}^nI(w^i>0)$ when the kernel $K_\epsilon$ has compact support.
Alternative definitions of the effective number of simulations, such as
$c\sum_{i=1}^nw^i$ for some $c>0$, can be on an arbitrary scale, since
the least squares regression solution is insensitive to the scale of
the weights. For any fixed value of $c$, the value of $c\sum
_{i=1}^nw^i$ will decrease as $p=\dim(s)$ increases so that it will
artificially favor larger numbers of (even uninformative) summary statistics.
Our definition of $\tilde{n}$ guarantees that the AIC scores are
comparable for different subsets of summary statistics. A downside is
that this definition of $\tilde{n}$ is only suitable for kernels,
$K_\epsilon$, with a compact support.

In equation (\ref{eqAIC}), $\hat{\sigma}_j^2$ is defined as the
weighted mean of squared residuals for the regression of $\theta_j$ and
is given by
\[
\hat{\sigma}^2_j = \frac{\sum_{i=1}^n w^i[\theta^i_j -\hat
{m}_j(s^i)]^2}{\sum_{i=1}^n w^i},
\]
where $\theta^i_j$ is the $j$th component of $\theta^i$ and $\hat
{m}_j(s)$ denotes the estimate of the mean function $m_j(s)=\mathbb
{E}[\theta_j|s]$.
For small sample sizes, the corrected AIC, the so-called AICc, is given
by replacing $d$ in (\ref{eqAIC}) by $d(d+1)/(\tilde{n}-d-1)$\vadjust{\goodbreak}
(\cite{HurvichTsai89}).
In the same manner the BIC can be defined as
%
%e16 #&#
\begin{equation}
\label{eqBIC} \mbox{BIC}=\tilde{n}\log{\prod_{j=1}^q
\hat{\sigma}_j^2} +d \log\tilde{n}.
\end{equation}
Alternative penalty terms involving the hat matrix of the regression
could also be used in the above (e.g.,
\cite{HurvichEtAl98}; \cite{Irizarry01}; \cite{KonishiEtAl04}).

It is instructive to note that in using the linear regression
adjustment (\ref{eqcorrec}), the above information criteria may be
expressed as
\[
\mbox{xIC}=\tilde{n}\log{\prod_{j=1}^q \operatorname{Var} \bigl(\theta_j^{*}\bigr)} +\mbox
{penalty term},
\]
where $\theta^{*}_j$ is the $j$th element of the regression adjusted
vector $\theta^{*}=(\theta^{*}_1,\ldots,\theta^{*}_q)$.
As such,\vspace*{1pt} up to the penalty terms, both AIC and BIC seek the
combination of summary statistics that minimizes the product of the
marginal variances of the adjusted posterior sample. Similarly to the
entropy criterion of \citet{NunesBalding10} (see Section
\ref{secentropy}), these information criterion will select those
summary statistics that maximize the precision of the posterior
distribution, $p_{\mathrm{ABC}}(\theta|s_{\mathrm{obs}})$. However, unlike
\citet{NunesBalding10}, this precision is\break traded off by a penalty
for model complexity.

A rationale for the construction of AIC and BIC in this manner is that
the summary statistics that should be included within an ABC analysis
are those which are good predictors of $\theta$. However, an obvious
requirement for AIC or BIC to identify an informative statistic is that
the statistic varies (with $\theta$) within the local range of the
regression model. If a statistic is informative outside of this range,
but uninformative within it, it will not be identified as informative
under these criteria.

%%%%%%%%%%%%%%%%%%%%%%%%%%%%%%%%%%%%%%%%%%%%%%%%%
%%%%%%%%%%%%%%%%%%%%%%%%%%%%%%%%%%%%%%%%%%%%%%%%%%
%s3.2 #&#
\subsection{Regularization via Ridge Regression}
%%%%%%%%%%%%%%%%%%%%%%%%%%%%%%%%%%%%%%%%%%%%%%%%%%
%%%%%%%%%%%%%%%%%%%%%%%%%%%%%%%%%%%%%%%%%%%%%%%%%
\label{sectionridge}

As described in Section~\ref{sectionregression-adjustment}, the
local-linear regression adjustment of \citet{BeaumontEtAl02} fits
the linear model
\[
\theta^i=\alpha+\beta^\top s^{i}+e^i
\]
based on the prior predictive samples $(\theta^i,s^i)\sim\break p(s|\theta
)p(\theta)$ and with regression weights given by $w^i=K_\epsilon(\|
s^i-s_{\mathrm{obs}}\|)$.
(As before, we describe the case where $\theta$ is univariate for
notational simplicity and clarity of exposition, but the approach
outlined below can be readily implemented for each component of a
multivariate $\theta$.)
However, in fitting the model by minimizing the weighted least squares
criteria,\vspace*{1pt}\vadjust{\goodbreak} $\sum_{i=1}^nw^i\|\alpha-\beta^\top{s^i}-\theta^i\|^2$, there
is a risk of over-adjustment by adjusting the parameter values via (\ref
{eqcorrec}) in the direction of uninformative summary statistics.

To avoid this, implicit dimension reduction within the regression
framework can be performed by alternatively minimizing the regularized
weighted sum of squares (\cite{HoerlKennard70})
%
%e17 #&#
\begin{equation}
\label{eqridge} \sum_{i=1}^n
w^i\bigl\|\theta^i - \bigl(\alpha+\beta^\top
s^{i}\bigr) \bigr\|^2 +\lambda\| \beta\|^2
\end{equation}
with regularization parameter $\lambda>0$.
As with the regularization component within the neural network model of
\citet{BlumFrancois10} (Section~\ref{sectionneural-networks}),
with ridge regression the risk of over-adjustment is reduced because
the regression coefficients, $\beta$, are shrunk toward zero by
imposing a penalty on their magnitudes. Note that while we consider
ridge regression here, a~number of alternative regularization
procedures could be implemented, such as the Lasso method.

An additional\vspace*{1.5pt} advantage of ridge regression is that standard least
squares estimates, $(\hat{\alpha}_{\mathrm{LS}},\hat{\beta}_{\mathrm{LS}})
^\top=(X^\top\cdot WX)^{-1} X^\top W \Theta$, are not guaranteed to have a unique
solution. Here
$X$ is a $n\times(p+1)$ design matrix given in equation (\ref{eqdesign}),
$\Theta=(\theta^1,\ldots,\theta^n)$ is the $n\times1$ column vector of
sampled $\theta^i$, and
$W=\mbox{diag}(w^1,\ldots,w^n)$ is an $n\times n$ diagonal matrix of weights.
The lack of a unique solution can arise through multicolinearity of the
summary statistics, which can result in singularity of the matrix
$X^\top WX$.
In contrast, minimization of the regularized weighted sum of squares
(\ref{eqridge}) always has a unique solution, provided\vspace*{1pt} that $\lambda
>0$. This solution is given by
$(\hat{\alpha}_{\mathrm{ridge}},\hat{\beta}_{\mathrm{ridge}})^\top=(X^\top WX +
\lambda I_p)^{-1} X^\top W \Theta$, where $I_p$ denotes the $p\times p$
identity matrix.
There are several approaches for dealing with the regularization
parameter $\lambda$, including cross-validation and generalized
cross-validation to identify an optimal value of $\lambda$ (\cite
{GolubEtAl79}), as well as averaging the regularized estimates $(\hat
{\alpha}_{\mathrm{ridge}},\hat{\beta}_{\mathrm{ridge}})^\top$ obtained for
different values of $\lambda$ (\cite{TaniguchiAndTresp97}).

%%%%%%%%%%%%%%%%%%%%%%%%%%%%%%%%
%%%%%%%%%%%%%%%%%%%%%%%%%%%%%%%%%
%s4 #&#
\section{A Comparative Analysis}
%%%%%%%%%%%%%%%%%%%%%%%%%%%%%%%%%
%%%%%%%%%%%%%%%%%%%%%%%%%%%%%%%%
\label{sectionexamples}

We now provide a comparative analysis of the previously described
methods of dimension reduction within the context of three previously
studied analyses in the ABC literature.
Specifically, this includes the analysis of a coalescent model with
recombination (\cite{JoyceMarjoram08}), an evaluation of the evolutionary
fitness cost of mutation\vadjust{\goodbreak} in drug-resistant tuberculosis (\cite
{LucianiEtAl09}) and an assessment of the number and size-distribution
of particle inclusions in the production of clean steels (\cite
{BortotEtAl07}).

Each analysis is based on $n=1\mbox{,}000\mbox{,}000$ simulations where the parameter
$\theta$ is drawn from the prior distribution $p(\theta)$. The
performance of each method is measured through the
$\overline{\mbox{RSSE}}$ criterion (\ref{eqnmrsse}) following
\citet{NunesBalding10}, based on the same randomly selected subset
of $n^*=100$ samples $(\theta^i,\break s^i)=(\theta_{\mathrm{true}},s_{\mathrm{obs}})$ as
``observed'' data sets. When evaluating the RSSE error measure of
equation (\ref{eqnrsse}), we give a weight $w^i=1$ for the accepted
simulations and a weight of 0 otherwise. As the value of the RSSE
(\ref{eqnrsse}) depends on the scale of each parameter, we standardize
the parameters in each example by dividing the parameter values by the
standard deviation obtained from the $n=1\mbox{,}000\mbox{,}000$ simulations (with
the exception of the first example, where the parameters are on similar
scales).
For comparative ease, and to provide a performance baseline for each
example, all $\overline{\mbox{RSSE}}$ results are presented as relative
to the $\overline{\mbox{RSSE}}$ obtained when using the maximal vector
of summary statistics and no regression adjustment. In this manner, a
relative $\overline{\mbox{RSSE}}$ of $x/\mbox{$-x$}$ denotes an $x$\%
worsening/improvement over the baseline score.

Within each ABC analysis, we use Euclidean distance within an
Epanechnikov kernel $K_\epsilon(\|s-s_{\mathrm{obs}}\|)$. The Euclidean
distances are computed after standardizing the summary statistics with
a robust estimate of the standard deviation (the mean absolute
deviation). The kernel scale parameter, $\epsilon$, is determined as
the value at which exactly 1\% of the simulations, $(\theta^i,s^i)$,
have nonzero weight. This yields exactly $\tilde{n}=10\mbox{,}000$ simulations
that form the final sample from each posterior.
To perform the method of \citet{fernhead+p11}, a randomly chosen
10\% of the $n$ simulations were used to fit the regression model that
determines the choice of summary statistics, with the remaining 90\%
used for the ABC analysis. The final ABC sample size $\tilde{n}=10\mbox{,}000$
was kept equal to the other methods by slightly adjusting the scale
parameter, $\epsilon$.
In addition, for the method of \citet{fernhead+p11}, following
exploratory analyses, the regression model (\ref{fp11}) was fitted
using $f(s) = (s,s^2,s^3,s^4)$ for examples 1 and 2 (as described in
Section~\ref{sectionFP11}) and using $f(s)=(\log s, [\log s]^2,[\log
s]^3,[\log s]^4)$ for example 3, always resulting in $4 \times p$
independent variables in the regression model of equation (\ref{fp11}).

When using neural networks or ridge regression to estimate the
conditional mean and variance, $m(s)$ and $\sigma^2(s)$, we take the
pointwise median of the estimated functions obtained with the
regularization parameters $\lambda=10^{-3},10^{-2}$ and $10^{-1}$.
These values of $\lambda$ assume that the summary statistics and the
parameters have been standardized before fitting the regression
function (\cite{ripley94}).
However, because the optimization procedure for neural networks (the
\texttt{R} function \texttt{nnet}) only finds local optima, in this case we
take the pointwise median of ten estimated functions, with each
optimization initialized from a different random starting point, and
randomly choosing the regularization parameter with equal probability
from the above values (see Tanigu\-chi and Tresp (\citeyear{TaniguchiAndTresp97})).

%%%%%%%%%%%%%%%%%%%%%%%%
%%%%%%%%%%%%%%%%%%%%%%%%%
%s4.1 #&#
\subsection{Example 1: A Coalescent Analysis}
%%%%%%%%%%%%%%%%%%%%%%%%%
%%%%%%%%%%%%%%%%%%%%%%%%

This model was previously considered by \citet{JoyceMarjoram08}
and \citet{NunesBalding10}, each while proposing their respective
ABC dimension reduction strategies (see Sections
\ref{sectionJoyceMarjoram} and~\ref{secentropy}). The analysis focuses on
joint estimation of the scaled mutation rate, $\tilde{\theta}$, and the
scaled recombination rate, $\rho$, in a coalescent model with
recombination (\cite{Nordborg07}).
Under this model, 5001 base pair DNA sequences for $50$ individuals
are generated from the coalescent model, with recombination, under the
infi\-nite-sites mutation model, using the software \texttt{ms}\break (\cite{Hudson02}).
The initial summary statistics, $s=(s_1,\ldots,s_6)$, are the number of
segregating sites ($s_1$), the pairwise mean number of nucleotidic
differences ($s_2$), the mean $R^2$ across pairs separated by $<$10\%
of the simulated genomic regions ($s_3$), the number of distinct
haplotypes ($s_4$), the frequency of the most common haplotype ($s_5$)
and the number of singleton haplotypes $(s_6)$.

We first examine the performance of ABC without using dimension
reduction techniques. For different parameter combinations, $\tilde
{\theta}, \rho$ and $(\tilde{\theta},\rho)$, we compute the relative
$\overline{\mbox{RSSE}}$ obtained with a single optimal summary
statistic and the relative $\overline{\mbox{RSSE}}$ obtained when using
all six population genetic\vspace*{1pt} statistics ($s_1$--$s_6$) (Table~\ref
{tabex1-2}). When estimating $\tilde{\theta}$ only, we find that using
only the number of segregating sites ($s_1$) provides lower relative
%
%t1 #&#
\begin{table*}
\tablewidth=370pt
\caption{Relative $\overline{\mbox{RSSE}}$ for examples 1 and 2.
The leftmost column shows the minimal $\overline{\mbox{RSSE}}$ when
considering only one summary statistic (with no regression adjustment).
Rightmost columns show relative $\overline{\mbox{RSSE}}$ using all
summary statistics under no, homoscedastic and heteroscedastic
regression adjustment. All $\overline{\mbox{RSSE}}$ are relative to the
$\overline{\mbox{RSSE}}$ obtained when using no regression adjustment
with all summary statistics. The score of the best method in each
analysis (row) is emphasised in boldface}
\label{tabex1-2}
%
%@{\hspace*{12pt}}c@{\hspace*{-3pt}}k{2.0}@{\hspace*{-3pt}}k{2.0}@{}}
%&& \multicolumn{1}{@{\hspace*{-1pt}}c}{\multirow{2}{76pt}[-6.5pt]{\centering{\textbf{One optimal statistic (no adj.)}}}}
%& \multicolumn{3}{@{}c@{}}{\textbf{All summary statistics}}\\
%&& & \multicolumn{1}{@{}c@{\hspace*{-3pt}}}{\textbf{No adj.}}
%& \multicolumn{1}{c}{\textbf{Homo adj.}\hspace*{-1pt}}
%& \multicolumn{1}{@{\hspace*{-1pt}}c@{}}{\textbf{Hetero adj.}}\\
%Example 1 & $\tilde{\theta}$ & \mbox{$\bolds{-7}$},\ \mbox{($s_1$)} & 0 & -3,& -3, \\
%& $\rho$ & 9,\ \mbox{($s_5$)} & 0 & \mbox{$\bolds{-5}$}, & -4,\\
%&$(\tilde{\theta},\rho)$ & 7,\ \mbox{($s_1$)} & 0 & 0, & \mbox{$\bolds{-7}$},\\
%[6pt]
%Example 2 &$\alpha$ & 6,
%& 0 & \mbox{$\bolds{-3}$}, & \mbox{$\bolds{-3}$}, \\
%&$c$ & \mbox{$\bolds{-7}$},
%& 0 & -5, &-5,\\
%&$\rho$ & \mbox{$\bolds{-9}$},
%& 0 & -8, &-8,\\
%&$\mu$ &\mbox{$\bolds{-14}$},
%& 0&-5,&-6,\\
%&$(\alpha,c,\rho,\mu)$ & 5,
%& 0 &\mbox{$\bolds{-4}$},&\mbox{$\bolds{-4}$},\\
%
\begin{tabular*}{\tablewidth}{@{\extracolsep{4in minus 4in}}l@{\hspace*{16pt}}ck{4.4}ck{2.0}k{2.0}@{}}
\hline
&& \multicolumn{1}{c}{\multirow{2}{76pt}[-9.5pt]{\centering{\textbf{One optimal statistic (no adj.)}}}}
& \multicolumn{3}{c@{}}{\textbf{All summary statistics}}\\[-2pt]
&&& \multicolumn{3}{l@{}}{\rule{151pt}{1pt}}\\
&& & \multicolumn{1}{c}{\textbf{No adj.}}
& \multicolumn{1}{c}{\textbf{Homo adj.}}
& \multicolumn{1}{c@{}}{\textbf{Hetero adj.}}\\
\hline
Example 1 & $\tilde{\theta}$ & \mbox{$\bolds{-7}$},\ \mbox{($s_1$)} & 0 & -3,& -3, \\
& $\rho$ & 9,\ \mbox{($s_5$)} & 0 & \mbox{$\bolds{-5}$}, & -4,\\
&$(\tilde{\theta},\rho)$ & 7,\ \mbox{($s_1$)} & 0 & 0, & \mbox{$\bolds{-7}$},\\
[6pt]
Example 2 &$\alpha$ & 6,
& 0 & \mbox{$\bolds{-3}$}, & \mbox{$\bolds{-3}$}, \\
&$c$ & \mbox{$\bolds{-7}$},
& 0 & -5, &-5,\\
&$\rho$ & \mbox{$\bolds{-9}$},
& 0 & -8, &-8,\\
&$\mu$ &\mbox{$\bolds{-14}$},
& 0&-5,&-6,\\
&$(\alpha,c,\rho,\mu)$ & 5,
& 0 &\mbox{$\bolds{-4}$},&\mbox{$\bolds{-4}$},\\
\hline
\end{tabular*}
\end{table*}
$\overline{\mbox{RSSE}}$ than when including all 6 summary statistics
even when performing regression adjustment. For all other parameter
combinations, using a single statistic produces substantially worse
than the rejection algorithm with all summary statistics. For all
inferences [i.e., of $\tilde{\theta}$, $\rho$ and $(\tilde{\theta},\rho
)$ jointly], regression adjustments generally improve the inference
when using all six summary statistics, which is consistent with
previous results (\cite{NunesBalding10}). The only exception is when
jointly estimating $(\tilde{\theta},\rho)$, where homoscedastic linear
adjustment neither decreases nor increases the error obtained with the
pure rejection algorithm.

\begin{figure*}[b]

\includegraphics{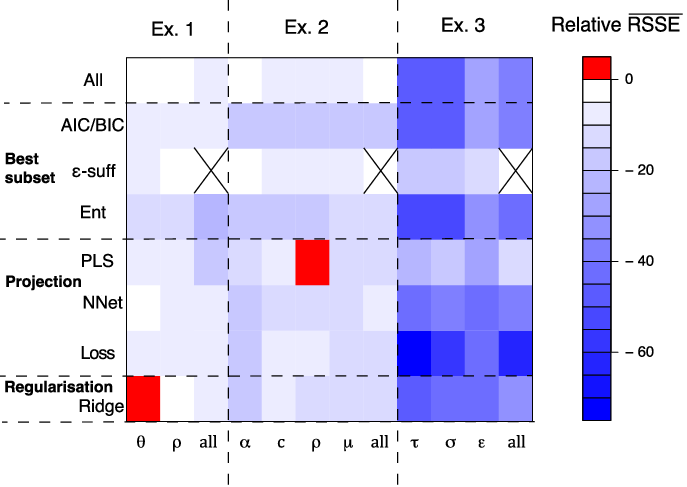}

\caption{Relative $\overline{\mbox{RSSE}}$ for the different methods of
dimension reduction in the three examples. All $\overline{\mbox{RSSE}}$
are relative to the $\overline{\mbox{RSSE}}$ obtained when using no
regression adjustment with all summary statistics. Methods of dimension
reduction include no dimension reduction (All), AIC/BIC, the
$\varepsilon$-sufficiency criterion ($\varepsilon$-suff), the two-stage
entropy procedure (Ent), partial least squares (PLS), neural networks
(NNet), minimum expected posterior loss (Loss) and ridge regression
(Ridge). The crosses correspond to situations for which there is no
result available.}
\label{figblue}
\end{figure*}

Next, we investigate the performance of each dimension reduction
technique. Table~\ref{tabex1-3} and Figure~\ref{figblue} show the
relative $\overline{\mbox{RSSE}}$ obtained under each dimension
reduction method for each parameter combination and under
heteroscedastic regression adjustment. For all three examples, more
complete tables that contain the results obtained with no regression
adjustment and homoscedastic adjustment can be found in the
supplementary information to this article (\cite{Bluetal}).

%t2 #&#
\begin{table*}
\caption{Relative $\overline{\mbox{RSSE}}$ for examples 1--3 for
different parameter combinations using each method of dimension
reduction,\break and under heteroscedastic regression adjustment.
Columns denote no dimension reduction (All), BIC, AIC,\break AICc, the
$\varepsilon$-sufficiency criterion ($\varepsilon$-suff), the two-stage
entropy procedure (Ent), partial least squares (PLS),\break neural networks
(NNet), minimum expected posterior loss (Loss) and ridge regression (Ridge).
All $\overline{\mbox{RSSE}}$ are\break relative to the $\overline{\mbox
{RSSE}}$ obtained when using no regression adjustment with all summary
statistics. The score of\break the best method in each analysis (row) is
emphasized in boldface}
\label{tabex1-3}
\begin{tabular*}{\tablewidth}{@{\extracolsep{4in minus 4in}}ld{3.0}d{3.0}d{3.0}d{3.0}d{3.0}
d{3.0}d{3.0}d{3.0}d{3.0}d{3.0}@{}}
\hline
& & \multicolumn{5}{c}{\hspace*{-4pt}\textbf{Best subset selection}}& \multicolumn
{3}{c}{\hspace*{-2pt}\textbf{Projection techniques}} & \multicolumn{1}{c@{}}{\textbf{Regularization}}\\
[-2pt]
&& \multicolumn{5}{l}{\hspace*{-2pt}\rule{183.5pt}{1pt}}
& \multicolumn{3}{l}{\hspace*{-2pt}\rule{118pt}{1pt}}
& \multicolumn{1}{l@{}}{\rule{57pt}{1pt}}\\
& \multicolumn{1}{c}{\hspace*{-3pt}\textbf{All}} & \multicolumn{1}{c}{\textbf{BIC}}
& \multicolumn{1}{c}{\textbf{AIC}} & \multicolumn{1}{c}{\textbf{AICc}}
& \multicolumn{1}{c}{$\bolds{\varepsilon}$\textbf{-suff}}
& \multicolumn{1}{c}{\textbf{Ent}} & \multicolumn{1}{c}{\textbf{PLS}} &
\multicolumn{1}{c}{\textbf{NNet}$\bolds{\tnote
{1}}$}& \multicolumn{1}{c}{\textbf{Loss}} &
\multicolumn{1}{c@{}}{\textbf{Ridge}$\bolds{\tnote{1}}$}\\
\hline
$\tilde{\theta}$ &-3 & -5 & -5 & -5 & -6 &\mbox{$\bolds{-11}$} &-6 & -4 & -7 & 1
\\
$\rho$ &-4 & -6 & -6 & -6 & 0 & \mbox{$\bolds{-12}$} &-7 & -8 & -7 & -3 \\
$(\tilde{\theta},\rho)$ &-7 & -7 & -7 & -7 & \multicolumn{1}{c}{--} & \mbox{$\bolds{-24}$}& -16 & -7 &
-6 & -6 \\
[6pt]
$\alpha$ &-3 & -15 & -15 & -15& 0 &\mbox{$\bolds{-17}$} &-13 &-15& \mbox{$\bolds{-17}$} &-15\\
$c$ & -5 & \mbox{$\bolds{-15}$} & \mbox{$\bolds{-15}$} & \mbox{$\bolds{-15}$}& -8 &\mbox{$\bolds{-15}$} &-8 &-12& -9
&-9\\
$\rho$ & -8 & \mbox{$\bolds{-16}$} & \mbox{$\bolds{-16}$} & \mbox{$\bolds{-16}$}& -8 &\mbox{$\bolds{-16}$} &1 &-12&
-9 &-10 \\
$\mu$ & -6 &\mbox{$\bolds{-18}$} & \mbox{$\bolds{-18}$} & \mbox{$\bolds{-18}$} & -8 &-13 &-10 &-13& -12
&-12 \\
$(\alpha,c,\rho,\mu)$ & -4 & \mbox{$\bolds{-19}$} & \mbox{$\bolds{-19}$} & \mbox{$\bolds{-19}$} & \multicolumn{1}{c}{--}
&-13 &-10 &-9& -12 &-11 \\
[6pt]
$\tau$ &-49 & -47 & -47 & -48 & -19 & -52 & -22 & \multicolumn{1}{c}{$-20\mbox{$/$}{-}42$}& \mbox{$\bolds{-75}$} &
\multicolumn{1}{c@{}}{$-48\mbox{$/$}{-}48$} \\
$\sigma$ &-45 & -46 & -47 & -46 & -15 & -50 & -15 & \multicolumn{1}{c}{$-21\mbox{$/$}{-}37$} & \mbox{$\bolds{-56}$}
&\multicolumn{1}{c@{}}{$-43\mbox{$/$}{-}43$} \\
$\xi$ & -27 & -29 & -29 & -28 & -13 & -32 & -28 & \multicolumn{1}{c}{$-7\mbox{$/$}{-}41$} & -41 &
\multicolumn{1}{c@{}}{$-26$\mbox{$/$}\mbox{$\bolds{-44}$}} \\
($\tau,\sigma,\xi$) & -39 &-39 & -40 & -39 & \multicolumn{1}{c}{--}& -42 & -11 & \multicolumn{1}{c}{$-4\mbox{$/$}{-}38$}
&\mbox{$\bolds{-60}$} &\multicolumn{1}{c@{}}{$-39\mbox{$/$}{-}32$}\\
\hline
\end{tabular*}
\tabnotetext[1]{1}{For the third example, the first value is found by
integrating out the regularization parameter, whereas the second one is
found by choosing an optimal regularization parameter with
cross-validation. In examples 1 and 2, integration over the
regularization parameter is
performed.}
\end{table*}

The performance achieved with AIC, AICc or BIC is comparable to (i.e.,
the same or slightly better than) the result obtained when including
all six population genetic statistics. When using the $\varepsilon
$-sufficiency criterion, we find that the performance is improved for
the inference on $\tilde{\theta}$ only. The only best subset selection
method for dimension reduction that substantially and uniformly
improves the performance of ABC posterior estimates is the
entropy-based approach.
For the projection techniques, all methods (partial least squares,
neural nets and minimum expected posterior loss) outperform the
adjustment method based on all six population genetics statistics, with
a large performance advantage for partial least squares when estimating
$(\tilde{\theta},\rho)$ jointly. By contrast, ridge regression provides
no improvement over the standard regression adjustment (the ``All'' column).

%f1 #&#

Based on these results, a loose performance ranking of the dimension
reduction methods can be obtained by computing, for each method, the
mean (relative) $\overline{\mbox{RSSE}}$ over all parameter
combinations $\tilde{\theta}$, $\rho$ and $(\tilde{\theta},\rho)$ using
the heteroscedastic adjustment.
The worst performers were ridge regression and the $\varepsilon
$-sufficiency criterion (with a mean relative $\overline{\mbox{RSSE}}$
of $-$3\%). These are followed by the standard regression adjustment
with all summary statistics ($-$5\%) and the AIC/BIC, neural nets and
the posterior loss method ($-$6\%). The best performing methods are
partial least squares ($-$10\%) and the two-stage entropy-based
procedure ($-$16\%).

%%%%%%%%%%%%%%%%%%%%%%%%%%%%%%%%%%%
%%%%%%%%%%%%%%%%%%%%%%%%%%%%%%%%%%%%
%s4.2 #&#
\subsection{Example 2: The Fitness Cost of Drug Resistant Tuberculosis}
%%%%%%%%%%%%%%%%%%%%%%%%%%%%%%%%%%%%
%%%%%%%%%%%%%%%%%%%%%%%%%%%%%%%%%%%

We now consider an example of Markov processes for epidemiological
modeling. If a pathogen, such as Mycobacterium tuberculosis, mutates to
gain an evolutionary advantage, such as antibiotic resistance, it is
biologically plausible that this mutation will come at a cost to the
pathogen's relative fitness. Based on a stochastic model to describe
the transmission and evolutionary dynamics of Mycobacterium
tuberculosis, and based on incidence and genotypic data of the
{IS6110} marker, \citet{LucianiEtAl09} estimated the posterior
distribution of the pathogen's transmission cost and relative fitness.
The model contained $q=4$ free parameters: the transmission rate,
$\alpha$, the transmission cost of drug resistant strains, $c$, the
rate of evolution of resistance, $\rho$, and the mutation rate of the
{IS6110} marker, $\mu$.

\citet{LucianiEtAl09} summarized information generated from the
stochastic model through $p=11$ summary statistics. These statistics
were expertly elicited as quantities that were expected to be
informative for one or more model parameters, and included the number
of distinct genotypes in the sample, gene diversity for sensitive and
resistant cases, the proportion of resistant cases and measures of the
degree of clustering of genotypes, etc. It is considered likely that
there is dependence and potentially replicate information within these
statistics.

As before, we examine the relative performance of the statistics
without using dimension reduction techniques.
Table~\ref{tabex1-2} shows that for the univariate analysis of $c$,
$\rho$ or $\mu$, performing rejection sampling ABC with a single,
well-chosen summary statistic can provide an improved performance over
a similar analysis using all 11 summary statistics, under any form of
regression adjustment.
In particular, the proportion of isolates that are drug resistant is
the individual statistic which is most informative to estimate $c$
(with a relative $\overline{\mbox{RSSE}}$ of $-$7\%) and $\rho$ ($-$9\%).
For the marker mutation rate, $\mu$, the most informative statistic is
the number of distinct genotypes, with a relative $\overline{\mbox
{RSSE}}$ of $-$14\%. Conversely, an analysis using all summary statistics
with a regression adjustment offers the best inferential performance
for $\alpha$ alone, or for $(\alpha,c,\rho,\mu)$.
These results provide support for recent arguments in favor of
``marginal regression adjustments'' (\cite{nott+fms11}), whereby the
univariate marginal distributions of a full multivariate ABC analysis
are replaced by separately estimated marginal distributions using only
statistics relevant for each parameter. Here, more precisely estimated
margins can improve the accuracy of the multivariate posterior sample,
beyond the initial analysis.

The performance results of each dimension reduction method are shown in
Table~\ref{tabex1-3} and Figure~\ref{figblue}.
In contrast with the previous example, here the use of the AIC/BIC
criteria can substantially decrease posterior errors. For example,
compared to the linear adjustment of all 11 parameters, which produces
a mean relative $\overline{\mbox{RSSE}}$ between $-$3\% and $-$8\%
depending on the parameter (Table~\ref{tabex1-3}), using the AIC/\break BIC
criteria results in a relative $\overline{\mbox{RSSE}}$ of between
$-$15\% and $-$19\%.
The $\varepsilon$-sufficiency criterion produces more equivocal
results, however, as the error is sometimes increased with respect to
baseline performance (e.g., $+6\%$ when estimating $\alpha$ with
homoscedastic adjustment) and sometimes reduced (e.g., $-$8\% for $c$,
$\rho$ and $\theta$ with heteroscedastic adjustment).
As with the previous \mbox{example}, the entropy criterion provides a clear
improvement to the ABC posterior, and this improvement is almost
comparable to that produced by AIC/BIC.
Finally, the projection and regularization methods mostly all provide
comparable and substantive improvements compared to the baseline error,
with only partial least squares producing more equivocal results (e.g.,
$+1\%$ when estimating $\rho$).

Based on these results, the loose performance ranking of the dimension
reduction methods determines the worst performers to be the standard
least squares regression adjustment (with a mean relative $\overline
{\mbox{RSSE}}$ of $-$5\%), the $\varepsilon$-sufficiency approach ($-$6\%)
and partial least squares ($-$8\%).
These are followed by ridge regression ($-$11\%), neural networks and
the posterior loss method ($-$12\%). The best performing methods for
this analysis are the two-stage entropy-based procedure ($-$15\%) and
the AIC/BIC criteria ($-$17\%).

\begin{figure*}

\includegraphics{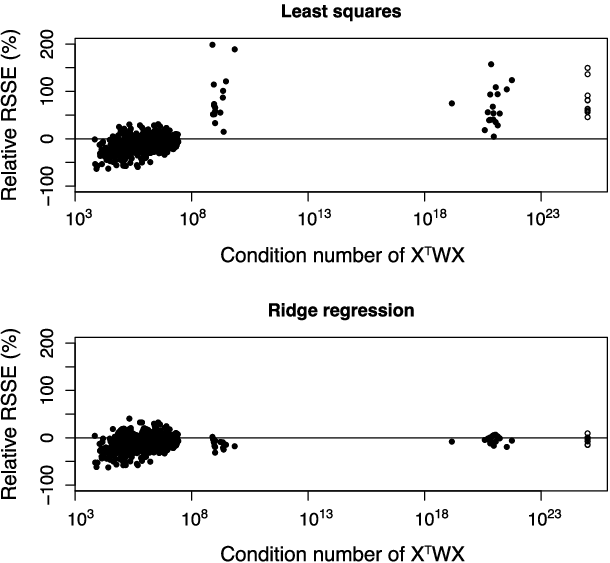}

\caption{Scatterplots of relative RSSE versus the condition number of
the matrix $X^\top WX$ for linear least squares (top) and ridge
(bottom) regression adjustments. Points are based on joint inference
for $(\alpha,c,\rho,\mu)$ in example 2 using 1000 randomly selected
vectors of summary statistics, $s^i$, as ``observed'' data. When the
minimum eigenvalue, $\lambda_{\mathrm{min}}$, is zero,\vspace*{1pt} the (infinite)
condition number is arbitrarily set to be $10^{25}$ for visual clarity
(open circles on the scatterplot).}
\label{figmulticollinearity}\vspace*{-3pt}
\end{figure*}

In this example, it is interesting to compare the performance of the
standard linear regression adjustment of all 11 summary statistics
(mean relative $\overline{\mbox{RSSE}}$ of $-$5\%) with that of the
ridge regression equivalent (mean relative $\overline{\mbox{RSSE}}$ of
$-$11\%). The increase in performance with ridge regression may be
attributed to its more robust handling of multicolinearity of the
summary statistics than under the standard regression adjustment.
To see this,\vspace*{1pt} Figure~\ref{figmulticollinearity} illustrates the
relationship between the relative $\overline{\mbox{RSSE}}$ (again,
relative to using all summary statistics\vspace*{1pt} and no regression adjustment)
and the condition number of the matrix $X^\top WX$, for both the
standard regression (top panel) and ridge regression (bottom panel)
adjustments based on inference for $(\alpha, c, \rho, \mu)$. The
condition number of $X^\top WX$ is given by
$\kappa=\sqrt{\lambda_{\mathrm{max}}/\lambda_{\mathrm{min}}}$, where
$\lambda_{\mathrm{max}}$ and $\lambda_{\mathrm{min}}$ are the largest and
smallest eigenvalues of $X^\top WX$. Extremely large condition numbers
are evidence for multicolinearity.

%f2 #&#

Figure~\ref{figmulticollinearity} demonstrates that for large values
of the condition number (e.g., for $\kappa>10^8$), the
least-squares-based regression adjustment clearly performs very poorly.
The region of $\kappa>10^8$ corresponds to almost 5\% of all
simulations,\vadjust{\goodbreak} and for these cases the relative error is hugely increased
(w.r.t. rejection) to anywhere between 5\% and 200\%. In contrast,\vspace*{1pt} for
ridge regression, the relative errors corresponding to $\kappa>10^8$
are not larger than the errors obtained for nonextreme condition numbers.
This analysis clearly illustrates that, unlike ridge regression, the
standard least squares regression adjustment can perform particularly
poorly when there is multicolinearity between the summary statistics.

In terms of the original analysis of \citet{LucianiEtAl09} which
used all eleven summary statistics with no regression adjustment
(although with a very low value for $\epsilon$), the above results
indicate that a more efficient analysis may have been achieved by using
a suitable dimension reduction technique.

%%%%%%%%%%%%%%%%%%%%%%%%%%%%%%%%%%%%
%%%%%%%%%%%%%%%%%%%%%%%%%%%%%%%%%%%%%
%s4.3 #&#
\subsection{Example 3: Quality Control in the Production of Clean Steels}
%%%%%%%%%%%%%%%%%%%%%%%%%%%%%%%%%%%%%
%%%%%%%%%%%%%%%%%%%%%%%%%%%%%%%%%%%%
\label{sectionexample3}

Our final example concerns the statistical modeling of extreme values.
In the production of clean steels, the occurrence of microscopic
imperfections (termed \textit{inclusions}) is unavoidable. The strength of
a clean steel block is largely dependent on the size of the largest
inclusion. \citet{BortotEtAl07} considered\vadjust{\goodbreak} an extreme value twist
on the standard stereological problem (e.g., \cite{baddeley+j04}),
whereby inference is required on the size and number of 3-dimensional
inclusions, based on data obtained from those inclusions that intersect
with a 2-dimensional slice. The model assumes a Poisson point process
of inclusion locations with rate parameter $\tau>0$ and that the
distribution of inclusion size exceedances above a measurement
threshold of $5\mu$m are drawn from a generalized Pareto distribution
with scale and shape parameters $\sigma>0$ and $\xi$, following
standard extreme value theory arguments (e.g., \cite{coles01}).

The observed data consist of 112 cross-sectional inclusion diameters
measured above $5\mu$m. The summary statistics thereby comprise 112
equally spaced quantiles of the cross-sectional diameters, in addition
to the number of inclusions observed, yielding $p=113$ summary
statistics in total. The ordering of the summary statistics creates
strong dependences between them, a~fact which can be exploited by
dimension reduction techniques. \citet{BortotEtAl07} considered
two models based on spherical or ellipsoidal shaped inclusions. Our
analysis here focuses on the ellipsoidal model.\vadjust{\goodbreak}

By construction, the large number ($2^{113}$) of possible combinations
of summary statistics means that the best subset selection methods are
strictly not practicable for this analysis, unless the number of
summary statistics is reduced further a priori. In order to facilitate
at least some comparison with the other dimension reduction approaches,
for the best subset selection methods \textit{only}, we consider six
candidate subsets. Each subset consists of the number of observed
inclusions in addition to 5, 10, 20, 50, 75 or 112 empirical quantiles
of the inclusion size exceedances (the latter corresponds to the
complete set of summary statistics). Due to the extreme value nature of
this analysis, the parameter estimates are likely to be more sensitive
to the precise values of the larger quantiles. As such, rather than
using equally spaced quantiles, we use a scheme which favors quantiles
closer to the maximum inclusion and we always include the maximum inclusion.

%
%t3 #&#
\begin{table*}
\tabcolsep=4pt
\caption{Summary of the main features of the different methods
of dimension reduction for ABC}
\label{tabsummary}
{\fontsize{8.5pt}{\baselineskip}\selectfont{\begin{tabular*}{\tablewidth}{@{\extracolsep{\fill}}lccccc@{}}
\hline
\textbf{Class} & \textbf{Method} & \textbf{Hyper-parameter}
& \textbf{Choice of hyper-parameter} &
\textbf{Computational burden} \\
\hline
Best subset selection & AIC/BIC & None & -- & Substantial/greedy alg. \\
& $\varepsilon$-suff& $T(\theta)$ & User choice & Substantial/greedy
alg. \\
& Ent & None & -- & Substantial/greedy alg. \\
[6pt]
Projection techniques & PLS & Number of PLS components, $k$ &
Cross-validation & Weak \\
& NNet & Regularization parameter, $\lambda$ & Integration or
cross-validation & Moderate (optimization algorithm) \\
&Loss & Choice of basis functions & BIC & Weak (closed-form solution) \\
[6pt]
Regularization & Ridge& Regularization parameter, $\lambda$ &
Integration or cross-validation & Weak (closed-form solution)\\
\hline
\end{tabular*}}}
\end{table*}
%

%f3 #&#
\begin{figure*}[b]

\includegraphics{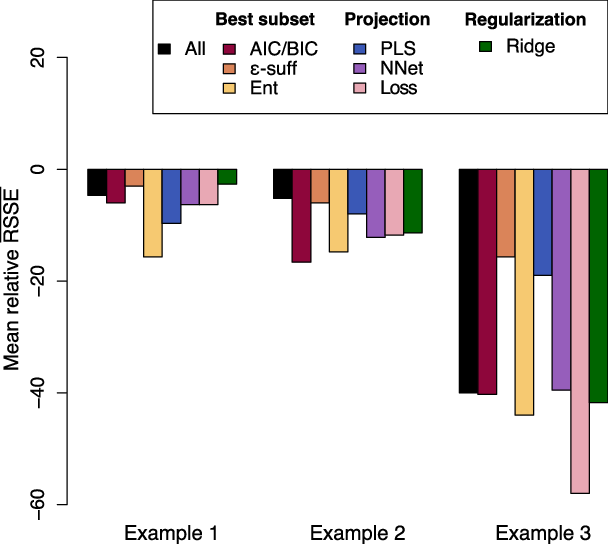}

\caption{Mean relative $\overline{\mbox{RSSE}}$ values using each
method of dimension reduction and for each example. Methods of
dimension reduction include no dimension reduction (All), AIC/BIC, the
$\varepsilon$-sufficiency criterion ($\varepsilon$-suff), the two-stage
entropy procedure (Ent), partial least squares (PLS), neural networks
(NNet), minimum expected posterior loss (Loss) and ridge regression
(Ridge). For examples 1 and 2, the results for ridge regression and
neural networks estimate $m(s)$ and $\sigma^2(s)$ have been obtained by
taking the pointwise median curve over varying values of the
regularization parameter; $\lambda=10^{-3},10^{-2}$ and $10^{-1}$ (see
introduction to Section~\protect\ref{sectionexamples}). For example 3, an
optimal value of $\lambda$ was chosen based on a cross-validation
procedure (see Section \protect\ref{sectionexample3}).}
\label{figbarplot}
\end{figure*}

The relative $\overline{\mbox{RSSE}}$ obtained under each dimension
reduction method is shown in Table~\ref{tabex1-3} and Figure \ref
{figblue}. In comparison to an analysis using all 113 summary
statistics and regression adjustment (the ``All'' column), the best
subset selection approaches do not in general offer any improvement.
While the entropy-based method provides a slight improvement, the
relative $\overline{\mbox{RSSE}}$ under the $\varepsilon$-sufficiency
criterion is substantially worse (along with partial least squares). Of
course, these results are limited to the few subsets of statistics
considered and it is possible that alternative subsets could perform
substantially better. However, it is computationally untenable to
evaluate this possibility based on exhaustive enumeration of all subsets.

When using neural networks to perform the regression adjustment based
on computing the pointwise median of the $m(s)$ and $\sigma^2(s)$
estimates, obtained using varying regularization parameter values (see
the introduction to Section~\ref{sectionexamples}), the relative
performance is quite poor (left-hand side $\overline{\mbox{RSSE}}$
values in Table~\ref{tabex1-3}). The mean relative $\overline{\mbox
{RSSE}}$ is $-$13\% for neural networks, compared to $-$40\% for
heteroscedastic least squares regression. As an alternative approach,
rather than averaging over the regularization parameter $\lambda$, we
rather choose the value of $\lambda\in\{10^{-3},10^{-2},10^{-1}\}$
that minimizes the leave-one-out error of $\theta$ [equation (\ref
{eqnPLSLOO})]. This approach considerably improves the performance of
the network (right-hand side $\overline{\mbox{RSSE}}$ values in Table
\ref{tabex1-3}) with the mean relative $\overline{\mbox{RSSE}}$
improving to the same level as for heteroscedastic regression.
Adopting the same procedure to determine the regularization parameter
within ridge regression, there is also a mean gain in performance from
$-$39\% to $-$42\%, although the joint parameter inference on $(\tau,
\sigma,\xi)$ actually performs worse under this alternative approach.
The variability in these results highlights the importance of making an
optimal choice of the regularization parameter for an ABC analysis.

The minimum expected posterior loss approach performs particularly well
here. This approach has also been shown to perform well in a similar
analysis: that of performing inference using quantiles of a large
number of independent draws from the (intractable) $g$-and-$k$
distribution (\cite{fernhead+p11}).

The loose performance ranking of each of the dimension reduction
methods finds that the worst performers are the $\varepsilon
$-sufficiency criterion (with a mean relative $\overline{\mbox{RSSE}}$
of $-$16\%) and partial least squares ($-$19\%). Neural networks and
AIC/BIC perform just as well as standard least squares regression
($-$40\%), ridge regression slightly outperforms standard regression
($-$42\%) and the entropy-based approach is a further slight
improvement at $-$44\%. The clear winner in this example is the
posterior loss approach with a mean relative $\overline{\mbox{RSSE}}$
of $-$58\%.

%%%%%%%%%%%%%%%%%%%%
%%%%%%%%%%%%%%%%%%%%%
%s5 #&#
\section{Discussion}
%%%%%%%%%%%%%%%%%%%%%
%%%%%%%%%%%%%%%%%%%%
\label{sectiondiscussion}

The process of dimension reduction is a critical and influential part
of any ABC analysis. In this article we have provided a comparative
review of the major dimension reduction approaches (and introduced two
new ones) in order to provide some guidance to the practitioner in
choosing the most appropriate technique for their own analysis. A
summary of the qualitative features of each dimension reduction method
is shown in Table~\ref{tabsummary}, and a comparison of the relative
performances of each method for each example is illustrated in Figure
\ref{figbarplot}. As with each individual example, we may compute an
overall performance ranking of the dimension reduction methods by
averaging the mean relative $\overline{\mbox{RSSE}}$ values over the
examples. Performing worse, on average, than a standard least squares
regression adjustment with no dimension reduction (with an overall mean
relative $\overline{\mbox{RSSE}}$ of $-$17\%) is the $\varepsilon
$-sufficiency technique ($-$8\%) and partial least squares ($-$12\%).
Performing better, on average, than standard least squares regression
is ridge regression and neural networks ($-$19\%) and AIC/BIC
($-$21\%). In this study, the top performers, on average, were the
entropy-based procedure and the minimum expected posterior loss
approach, with an overall mean relative $\overline{\mbox{RSSE}}$ of
$-$25\%.
It is worth emphasizing that the potential gains in performing a
regression adjustment alone (with all summary statistics and \textit{no}
dimension reduction) can be quite substantial. This suggests that
regression adjustment should be an integral part of the majority of ABC
analyses. Further gains in performance can then be obtained by
combining regression adjustment with dimension reduction procedures,
although in some cases (such as with the $\varepsilon$-sufficiency
technique and partial least squares) performance can sometimes worsen.

While being ranked in the top three, a clear disadvantage of the
entropy-based procedure and the AIC/BIC criteria is the quantity of
computation required. This primarily occurs as the best subset
selection procedures require evaluation of all $2^p$ potential models.
For examples 1 and 2, a greedy algorithm was able to find the optimum
solution in a reasonable time. This was not possible for example~3.
Additionally, in this \mbox{latter} case, for the subsets of summary
statistics considered, the performance obtained by implementing
computationally expensive methods of dimension reduction was barely an
improvement over the computationally cheap, least squares regression
adjustment. This raises the important point that the benefits of
performing potentially expensive forms of dimension reduction over,
say, the simple linear regression adjustment, should be evaluated prior
to their full implementation. We also note that the second stage of the
entropy-based method (Section~\ref{secentropy}) targets minimization
of (\ref{eqnmrsse}), the same error measure used in our comparative
analysis. As such, this approach is likely to be numerically favored in
our results.

The top ranked (ex aequo) minimum expected posterior loss approach
particularly outperforms\break other dimension reduction methods in the final
example (the production of clean steels). In such analyses, with large
numbers of summary statistics (here $p=113$), nonlinear methods such as
neural networks may become overparametrized, and simpler alternatives,
such as least squares or ridge regression adjustment, can work more effectively.
This is naturally explained through the usual bias-variance trade-off:
more complex regression models such as neural networks reduce the bias
of the estimate of $m(s)$ [and optionally $\sigma^2(s)$], but in doing
so the variance of the estimate is increased. This effect can be
especially acute for high-dimensional regression (\cite{geman1992neural}).

Our analyses indicate that the original least\break squares, linear
regression adjustment (Beaumont,\break Zhang and Balding (\citeyear{BeaumontEtAl02})) can sometimes perform
quite well, despite its simplicity. However, the presence of
multicolinearity between the summary statistics can cause severe
performance degradation, compared to not performing the regression
adjustment (see Figure~\ref{figmulticollinearity}). In such
situations, regularization procedures, such as ridge regression (e.g.,
example 2 and Figure~\ref{figmulticollinearity}) and projection
techniques, can be beneficial.

However, an important issue with regularization procedures, such as
neural networks and ridge regression, is the handling of the
regularization parameter, $\lambda$. The ``averaging'' procedure that was
used in the first two examples proved quite suboptimal in the third,
where a cross-validation procedure to select a single best parameter
value produced much improved results. This problem can be particularly
critical for neural networks with large numbers of summary statistics,
$p$, as the number of network weights is much larger than $p$, and,
accordingly, massive shrinkage of the weights (i.e., large values of
$\lambda$) is required to avoid overfitting.

The posterior loss approach produced the superior performance in the
third example. In general, a strong performance of this method can be
primarily attributed to two factors. First, in the presence of large
numbers of highly dependent summary statistics, the extra analysis
stage in determining the most appropriate regression model (\ref{fp11})
by choosing $f(s)$ through, for example, BIC diagnostics, affords the
opportunity to reduce the complexity of the regression in a simple and
relatively low-parameterized manner. This was not a primary contributor
in example 3, however, as the regression [equation (\ref{fp11})] was
directly performed on the full set of 113 statistics. Given the
benefits of using regularization methods in this setting, it is
possible that a ridge regression model would allow a more robust
estimate of the posterior mean (as a summary statistic) as part of this
process. Second, the posterior loss approach determines the number of
summary statistics to be equal to the number of posterior
quantities of interest---in this case, $q=3$ posterior parameter
means. This small number of derived summary statistics naturally allows
more precise posterior statements to be made, compared to dimension
reduction methods that produce a much larger number of equally
informative statistics. Of course, the dimension advantage here is
strongly related to the number of parameters ($q=3$) and summary
statistics ($p=113$) in this example.
However, it is not fully clear how any current methods of dimension
reduction for ABC would perform for substantially more challenging
analyses with considerably higher numbers of parameters and summary
statistics. This is because the curse of dimensionality in ABC (\cite
{Blum10}) has tended to restrict existing applications of ABC methods to
problems of moderate parameter dimension, although this may change in
the future.

What is very apparent from this study is that there is no single
``best'' method of dimension reduction for ABC. For example, while the
posterior loss and entropy-based methods were the best performers for
example 3, AIC and BIC were ranked first in the analysis of example 2,
and partial least squares outperformed the posterior loss approach in
example 1. A number of factors can affect the most suitable choice for
any given analysis. As discussed above, these can include the number of
initial summary statistics, the amount of dependence and
multicolinearity within the statistics, the computational overheads of
the dimension reduction method, the requirement to suitably determine
hyperparameters and sensitivity to potentially large numbers of
uninformative statistics.

One important point to understand is that all of the ABC analyses of
this review were performed using the rejection algorithm optionally
followed by some form of regression adjustment. While alternative,
potentially more efficient and accurate methods of ABC posterior
simulation exist, such as Markov chain Monte Carlo or sequential Monte
Carlo based samplers, the computational cost of separately implementing
such an algorithm $2^p$ times (in the case of best subset selection
methods) means that such dimension reduction methods can become rapidly
untenable, even for small $p$. The price of the benefit of using the
more computationally practical, fixed large number of samples is that
decisions on the dimension reduction of the summary statistics will be
made on potentially worse estimates of the posterior than those
available under superior sampling algorithms. As such, the final
derived summary statistics may in fact not be those which are most
appropriate for subsequent use in, for example, Markov chain Monte Carlo
or sequential Monte Carlo based algorithms.

However, this price is arguably a necessity. It is practically
important to evaluate the performance of any dimension reduction
procedure in a given analysis. Here we used a criterion [the $\overline
{\mbox{RSSE}}$ of equation (\ref{eqnmrsse})] that is based on a
leave-one-out procedure.
When using a fixed, large number of samples, evaluation of such a
performance diagnostic is entirely practicable, as no further model
simulations are required. This idea is also relevant to methods of
dimension reduction for model selection (\cite
{barnesetal11}; \cite{Marin12}) where a misclassification rate based on a
leave-one-out procedure can serve as a comparative criterion.

\section*{Acknowledgments}

S. A. Sisson is supported by the Australian Research Council through the
Discovery Project Scheme (DP1092805). M. G. B. Blum is supported by the
French National Research Agency (DATGEN project,
ANR-2010-JCJC-1607-01).

\begin{supplement}%[id=suppA]
\stitle{Supplement to ``A Comparative Review of Dimension Reduction Methods in
Approximate Bayes\-ian Computation''}
\slink[doi]{10.1214/12-STS406SUPP} %[doi,text={...}] - jei reikia
%suskaldyti doi
\sdatatype{.pdf}
\sfilename{sts406\_supp.pdf}
\sdescription{The supplement contains for each of the three examples a
comprehensive comparison of the errors obtained with the different
methods of dimension reduction.}
\end{supplement}

% imsref loaded by lrinkeviciute, 2012-10-01 15:45:41
% imsref loaded by lrinkeviciute, 2012-10-01 16:08:38
% imsref loaded by lrinkeviciute, 2012-10-02 08:29:42

\end{document}